\documentclass[11pt]{article}

\usepackage{jheppub}
\usepackage{amsmath,amssymb}
\usepackage{calc}

\usepackage{color}
\definecolor{darkgreen}{rgb}{0,0.5,0}
\definecolor{darkblue}{rgb}{0,0,0.7}
\definecolor{darkred}{rgb}{0.5,0,0.0}
\definecolor{darkorange}{rgb}{0.8,0.4,0.0}

\newcommand{\ie}{i.e.\ }
\newcommand{\eg}{e.g.\ }

\newcommand{\Ym}{Y$_{\text{m}}$}
\newcommand{\ycut}{y_{\text{cut}}}
\newcommand{\zcut}{z_{\text{cut}}}
\newcommand{\zetacut}{\zeta_{\text{cut}}}

\newcommand{\Rtrim}{R_{\text{trim}}}

\title{Improved jet substructure methods: Y-splitter and variants with grooming}

\author[a]{Mrinal Dasgupta,}
\author[b]{Alexander Powling,}
\author[c]{Lais Schunk,}
\author[c]{and Gregory Soyez}
\affiliation[a]{Lancaster-Manchester-Sheffield Consortium for Fundamental Physics, School of Physics
  \& Astronomy, University of Manchester, Manchester M13 9PL, United
  Kingdom}
\affiliation[b]{School of Physics
  \& Astronomy, University of Manchester, Manchester M13 9PL, United
  Kingdom}
\affiliation[c]{IPhT, CEA Saclay, CNRS UMR 3681, F-91191
  Gif-Sur-Yvette, France}

\emailAdd{mrinal.dasgupta@manchester.ac.uk}
\emailAdd{lais.sarem-schunk@cea.fr}
\emailAdd{gregory.soyez@cea.fr}


\abstract{It has recently been demonstrated with Monte Carlo studies
  that combining the well-known Y-splitter and trimming techniques
  gives rise to important gains in the signal significance achievable
  for boosted electroweak boson tagging at high $p_t$.  Here we carry
  out analytical calculations that explain these findings from first
  principles of QCD both for grooming via trimming and via the
  modified mass-drop tagger (mMDT). We also suggest modifications to
  Y-splitter itself, which result in great simplifications to the
  analytical results both for pure Y-splitter as well as its
  combination with general grooming methods. The modifications also
  lead to further performance gains, while making the results largely
  independent of choice of groomer. We discuss the implications of
  these findings in the broader context of optimal methods for boosted
  object studies at hadron colliders.}
\begin{document}

\maketitle
\section{Introduction}\label{sec:intro}

In recent years jet substructure studies have become of central
importance to new physics searches and LHC phenomenology involving
highly boosted particles (for reviews and further references see
Refs.~\cite{Abdesselam:2010pt, Altheimer:2012mn, Altheimer:2013yza}).
When one considers the decays of boosted particles at the LHC,
i.e. those with $p_t \gg M$, we encounter a situation where the decay
products are collimated and hence often reconstructed in a single
``fat'' jet rather than forming multiple resolved jets. The
substructure of that jet offers important clues as to its origin
i.e. whether it is a QCD jet or a jet initiated by e.g. an electroweak
boson, top quark or hypothetical new particles.

The role of jet substructure analyses in discriminating signal from
QCD background jets was first discussed in
Ref.~\cite{Seymour:1993mx}. Subsequently
Ref.~\cite{Butterworth:2002tt} developed the Y-splitter algorithm to
tag jets arising from the hadronic two-body decays of W bosons.
Somewhat more recently the power of jet substructure analyses for
discoveries at the LHC was clearly highlighted in
Ref.~\cite{Butterworth:2008iy} in the context of Higgs boson
searches. Following this article there has been enormous interest in
jet substructure methods and in exploiting the boosted particle regime
at the LHC and even beyond, at potential future machines
\cite{Bothmann:2016loj}.  Several new jet substructure algorithms and
techniques have been developed and validated in the past few years and
are now commonly used in LHC searches and phenomenology
\cite{Abdesselam:2010pt, Altheimer:2012mn,
  Altheimer:2013yza}. Furthermore, the importance of the
boosted regime increases for ongoing run-2 LHC studies due to the
increased access to higher transverse momenta i.e. to TeV-scale jets.

Another important development, in the context of jet substructure, has
been the development of analytical calculations from first principles
of QCD, for many of the more commonly used techniques. For example
such calculations have been performed for the (modified)
MassDropTagger (m)MDT \cite{Dasgupta:2013ihk} , pruning
\cite{Ellis:2009su,Ellis:2009me} and trimming \cite{Krohn:2009th} in
Refs.~\cite{Dasgupta:2013ihk,Dasgupta:2013via}. Analytical
calculations have also been performed for the SoftDrop method
\cite{Larkoski:2014wba,Frye:2016okc,Frye:2016aiz} and for radiation constraining
jet shapes \cite{Larkoski:2015kga,Dasgupta:2015lxh} based on the
N-subjettiness class of variables \cite{Thaler:2010tr} and energy
correlation functions (ECFs) \cite{Larkoski:2013eya}. These
calculations have enabled a much 
more detailed and robust
understanding of jet substructure methods than was possible with
purely numerical studies from Monte Carlo event generators. They have
enabled meaningful comparisons of the performance of tools over a wide
kinematic range and revealed both advantages of and flaws in several
standard techniques.  Additionally, analytical understanding has
directly led to the design of new and superior tools such as the mMDT
and Y-pruning \cite{Dasgupta:2013ihk} , followed by the SoftDrop class
of observables \cite{Larkoski:2014wba} inspired in part by the
properties of the mMDT.  The mMDT and SoftDrop methods both have
remarkable theoretical properties (such as freedom from non-global
logarithms \cite{DassalNG1}) and substantially eliminate
non-perturbative effects, which render them amenable to high precision
calculations in perturbative QCD
\cite{Frye:2016okc,Frye:2016aiz}. Moreover they have proved to be
invaluable tools in an experimental context and are seeing widespread
use in LHC searches and phenomenology (for examples of some recent
applications see
e.g. Refs.~\cite{CMS:2016pwo,CMS:2016jog,CMS:2016ehh}. )

In spite of all the progress mentioned above, some key questions
remain as far as the development of substructure techniques is
concerned. One such question is whether it is possible to use our
analytical insight to make further performance gains relative to the
existing substructure methods including various taggers, groomers and
jet shapes such as N-subjettiness. This could include either the
construction of new optimal tools or the use of judicious combinations
of existing methods, inspired by the physics insights that have
recently been obtained via analytics.  In Ref.~\cite{Dasgupta:2015yua}
an explicit example was provided of the latter situation. There it was
shown via Monte Carlo studies that combining the existing Y-splitter
technique with trimming led to significant gains in performance and
this combination strikingly outperformed standard taggers (mMDT,
pruning, trimming and Y-pruning) for both Higgs and W boson tagging
especially at high $p_t$. This is in contrast to Y-splitter alone
which, although it was one of the earliest substructure methods
invented, performs relatively poorly and hence has not seen extensive
use.\footnote{One instance of its use was provided by the ``ATLAS top
  tagger'' \cite{Brooijmans:2008} but this itself has not been used
  recently to our knowledge.}

Ref.~\cite{Dasgupta:2015yua} identified the main reasons for the
success of the Y-splitter and trimming combination. Firstly it was
observed that Y-splitter is an excellent method for suppressing the QCD
background.  The reason identified for this was the basic form
of the jet mass distribution for QCD jets tagged with Y-splitter:
\begin{equation}
\label{eq:basicform}
\frac{\rho}{\sigma} \frac{d\sigma}{d\rho} \approx \frac{C_F
  \alpha_s}{\pi} \left(\ln \frac{1}{y}-\frac{3}{4} \right) \exp \left
  [-\frac{C_F \alpha_s}{2\pi} \ln^2 \frac{1}{\rho} \right ], \, \,
\rho < y \, ,
\end{equation}
where $\rho$ is the normalised squared jet-mass,
$\rho \equiv \frac{m^2}{p_t^2 R^2}$ with $m$ the jet mass, $p_t$ the
transverse momentum and $R$ the jet radius. The parameter $y$ is the
value chosen for the $\ycut$ parameter of Y-splitter, which we will
define more precisely in the next section.  The result quoted above is
an all-orders resummed result in a fixed-coupling approximation and
valid to leading (double) logarithmic accuracy in the exponent. While
it has been written above for the case of quark jets, it is
straightforward to write a corresponding formula for gluon-initiated
jets. The result has the general form of a prefactor, involving at
most a logarithm in $y$, multiplying an exponential Sudakov
suppression factor which is identical to that obtained for the plain
jet mass. In contrast, for the plain jet mass the prefactor involves a
$\ln \rho$ instead of a $\ln y$ term. The replacement of $\ln \rho$ by
a more modest $\ln y$ term, while maintaining the exponential Sudakov
suppression, is the principal reason why background jets are strongly
suppressed by Y-splitter.\footnote{As noted in
  Ref.~\cite{Dasgupta:2015yua} an essentially similar form is also
  obtained for Y-pruning which also performs better than several other
  methods at high $p_t$.}  In Ref.~\cite{Dasgupta:2015yua},
Eq.~\eqref{eq:basicform} was simply quoted without derivation, while
in the present article we shall explicitly derive it in
section~\ref{sec:pure}.

The second key observation made in Ref.~\cite{Dasgupta:2015yua} was
that Y-splitter alone has a poor signal efficiency similar to that for
plain ungroomed jets. This is due to the fact that there is no jet
grooming subsequent to the basic tagging step in Y-splitter which
results in loss of mass resolution due to underlying event and ISR
effects. Hence, in spite of its excellent background rejection pure
Y-splitter suffers in comparison to other standard substructure
taggers in terms of performance.

Finally it was noted in Ref.~\cite{Dasgupta:2015yua} that the addition
of grooming (via trimming) to Y-splitter considerably alleviated the
problems with signal efficiency. While this could perhaps be
anticipated, it was also observed that the use of trimming did not
seem to crucially affect the background rejection of Y-splitter. This
more surprising finding made trimming a nice complementary tool to
Y-splitter as it cured the issues seen with signal jets while leaving
the desirable behaviour on background jets, as given in
Eq.~\eqref{eq:basicform}, essentially unaltered.

We remind the reader that analytical calculations for trimming itself
have been carried out in Ref.~\cite{Dasgupta:2013ihk}. They revealed
the presence of multiple transition points in the jet mass
distribution as well as potential undesirable bumps in the background,
in regions close to the signal masses i.e. at masses near the electroweak
scale for TeV scale jet transverse momenta. On the other hand when trimming
is used subsequent to Y-splitter the mass distribution still closely
resembles the well-behaved Y-splitter distribution, rather than the
mass spectrum for trimming.\footnote{As we also demonstrate later,
  using trimming prior to Y-splitter returns a mass-spectrum that
  closely resembles that for trimming. Hence grooming should
  generally be performed after tagging with Y-splitter.}

All of the above observations certainly call for an analytical
understanding. It is therefore of interest to firstly derive the
result for Y-splitter quoted in Eq.~\eqref{eq:basicform}. Following
this, one needs to understand the form of the jet mass spectrum when
trimming is applied subsequent to Y-splitter. Given the undesirable
features of trimming we alluded to before (even if they are not as
manifest in the present case) it is also of interest to consider what
happens when other groomers are used instead of trimming, like the
mMDT. Lastly, in order to obtain further gains or a more robust
tagger, one may also seek to make variations in the Y-splitter method
itself. These modifications should be such that the most essential
features of Eq.~\eqref{eq:basicform} are left intact but other less
relevant subleading and non-perturbative terms are either better
controlled theoretically or altogether eliminated. It is these
developments that we seek to make in the present article.

The layout of this article is as follows: in section 2 we perform
resummed calculations for the jet mass distribution for jets tagged with
Y-splitter. We first compute the resummed result at leading logarithmic
accuracy in $\rho$ and hence in the fixed-coupling limit recover
Eq.~\eqref{eq:basicform}. We also augment the resummed formula to
examine the effects of terms that are formally subleading in $\rho$
(i.e. at best  single-logarithmic in $\rho$) but enhanced by logarithms of $y$.

In section 3 we study Y-splitter with grooming. We examine the
structure of logarithmic enhancements that emerge both in fixed-order
studies (up to order $\alpha_s^2$) as well as at all orders. Here we
study both trimming and mMDT as groomers and hence shed light on the
key observation that grooming does not radically affect the background
suppression seen with pure Y-splitter. 

We stress that for all the techniques studied in
this paper, our all-orders results are formally valid to leading
logarithmic accuracy in $\rho$ in the resummed exponent.
Additionally, we also retain some subleading (single-logarithmic in
$\rho$) terms such as those arising from hard-collinear emissions. We
will refer to this throughout as the (modified) leading logarithmic
accuracy (LL) approximation. We find, as has also been noted in our
past work on other substructure methods~\cite{Dasgupta:2013ihk} and
jet shapes~\cite{Dasgupta:2015lxh}, that the modified leading
logarithmic calculations are sufficient to explain the main features
of Y-splitter and its combination with groomers.
Additionally in some cases we are further able to account for terms which are
double-logarithmic in general, i.e. when counting $\ln \rho$ and
$\ln y$ on the same footing. These results will be explicitly
specified by the ``LL+LL$_y$'' superscript. The additional LL$_y$ terms
are included in particular to provide an estimate for the size of
subleading corrections responsible for differences between the
variants of Y-splitter we will study here.

Section 4 is devoted to variants of the Y-splitter method. Here we
first consider Y-splitter defined with mass declustering (generalised
$k_t$ \cite{Cacciari:2011ma} with $p=1/2$) rather than the standard
declustering based on $k_t$ and comment on the implications of this
modification.  We also investigate, in this section, the effect of
replacing the $\ycut$ condition of Y-splitter with a $\zcut$ condition
like that used as the default in pruning and trimming and suggested as
an alternative for mMDT \cite{Dasgupta:2013ihk}. We further study the
effects of a gentle pre-grooming using SoftDrop on jets tagged by
Y-splitter.

Section 5 is devoted to a detailed study of non-perturbative effects using
Monte Carlo event generators. 

Finally, in section 6 we summarise our findings, draw conclusions and
provide suggestions for further investigation.

\section{Y-splitter calculation: QCD background}\label{sec:pure}
We shall provide below the calculation for the impact of the
Y-splitter algorithm on the QCD jet mass distribution. The Y-splitter
method involves declustering a jet using the $k_t$ distance between
constituents $i$ and $j$, defined as usual as
\cite{Catani:1991hj,Catani:1993hr,Ellis:1993tq}
\begin{equation}
d_{ij} = \mathrm{min} \left(p_{ti}^2,p_{tj}^2 \right)\theta_{ij}^2,
\end{equation}
where $p_{ti}$ and $p_{tj}$ are the transverse momenta of the two
particles and $\theta_{ij}^2=(y_i-y_j)^2+(\phi_i-\phi_j)^2$ their
angular separation in the rapidity-azimuth plane.\footnote{All our
  calculations throughout this paper also apply to $e^+e^-$ collisions
  where we use the $k_t$ distance defined as
\begin{equation}
d_{ij} = 2 \mathrm{min} \left(E_i^2,E_j^2 \right)
\left(1-\cos\theta_{ij} \right),
\end{equation}
where we use $E_i$, the particle energies, instead of their transverse
momentum wrt the beam direction.  }

One examines the value of $d_{ij}$ produced in the first step of
declustering and places a cut either directly on $d_{ij}$ which one
can take to be $ \sim M_W^2$ or on the ratio of $d_{ij}$ to the
squared jet mass, i.e. use $\ycut = d_{ij}/m_j^2 > y$. These cuts are designed
to retain more symmetric signal splittings (i.e. a genuine two-pronged
structure) while discriminating against QCD background.  We shall
study the latter variant here which was shown in Monte Carlo studies
to give excellent performance in rejecting QCD background jets
\cite{Dasgupta:2015yua}.

The quantity that we shall study throughout this paper is the jet mass
distribution for QCD jets that is obtained after the application of
Y-splitter as well as that obtained from a combination of Y-splitter
and grooming methods that we shall specify later. We will obtain
results for the quantity $\frac{\rho}{\sigma} \frac{d\sigma}{d\rho}$
where $\rho$ is the standard variable $\rho =\frac{m^2}{R^2 p_t^2}$,
with $m$ the jet mass, $p_t$ its transverse momentum with respect to
the beam and $R$ the jet radius.

\subsection{Leading-order calculation}\label{sec:pure-lo}
We start by computing the result for the jet mass distribution for
jets that are tagged by Y-splitter. In order to generate leading
logarithmic contributions it is sufficient to consider contributions
from soft and collinear gluon emissions from a hard parton.

Therefore at leading order in QCD (order $\alpha_s$) we have to
consider a jet made up of a hard quark or gluon and a single
accompanying soft and collinear gluon.  Here we shall explicitly
consider the case of quark jets to begin with, but it is trivial to
obtain the corresponding results for gluon initiated jets from the
ones we derive below.

Let us write the four-momenta of the particles as 
\begin{equation}
  p = p_t \left(1,1,0,0\right), \:
  k = \omega_t\left(\cosh y,\cos\phi,\sin\phi,\sinh y\right),
\end{equation}
where $p$ is the four-momentum of the hard quark, written in terms of
its transverse momentum $p_t$ wrt the beam and where without loss of
generality we can set its rapidity wrt the beam to zero. Likewise
$\omega_t$ is the transverse momentum of the emitted soft gluon, with
rapidity $y$ and azimuthal angle $\phi$. In the soft and collinear
limit we have $\omega_t \ll p_t$ and $\theta^2=(y^2+\phi^2) \ll 1$.

Let us first study the jet mass distribution with a cut on
$d_{ij}/m^2$, with $m$ being the jet mass. In the soft and collinear
approximation $d_{ij}= \omega_t^2 \theta^2$ while
$m^2 = \omega_t p_t \theta^2$ so that we cut on the quantity
$x=\omega_t/p_t$ \ie the transverse momentum fraction of the gluon, such
that $x>y$. The calculation for the jet mass distribution with this
cut is then simple to write down
\begin{equation}\label{eq:Ysplit-LO-start}
\frac{1}{\sigma}\frac{d\sigma}{d\rho}^{\mathrm{LO,soft-coll.}}
= \frac{C_F \alpha_s}{\pi} \int_0^1 \frac{dx}{x} \frac{d\theta^2}{\theta^2}
\,\delta \left(\rho -x \theta^2 \right)
\Theta \left(x>y \right),
\end{equation}
where we have taken a fixed-coupling approximation.\footnote{Strictly
  speaking, there are anyway no running-coupling corrections at pure
  leading-order accuracy.} 
In writing (\ref{eq:Ysplit-LO-start}), we
have implicitly normalised all angles to $R$ so that $\theta$ runs up
to $1$ (instead of up to $R$) and all $R$ dependence that arises at
our accuracy is incorporated into our definition of
$\rho=m^2/(p_tR)^2$.
We stress that (\ref{eq:Ysplit-LO-start}) is valid in the leading
logarithmic approximation where it is sufficient to include soft and
collinear gluons. We have also assumed that the jet radius $R$ is
small and systematically neglected powers of $R$.
Unless explicitly mentioned, we will use this convention throughout
the rest of the paper.
Note that Eq.~(\ref{eq:Ysplit-LO-start}) is written for quark
jets. One can easily extrapolate this, and the following formulae, to
gluon jets by replacing $C_F$ by $C_A$ and using the appropriate
splitting function.

We can easily integrate (\ref{eq:Ysplit-LO-start}) to
obtain
\begin{equation}
\label{eq:yslo}
\frac{\rho}{\sigma} \frac{d\sigma}{d\rho}^\mathrm{{LO,soft-coll.}} = \frac{C_F \alpha_s}{\pi} \left(\ln
  \frac{1}{y} \Theta\left(y>\rho \right) + \ln \frac {1}{\rho}  \Theta
  \left(\rho>y \right) \right).
\end{equation}

The result above is identical to previous results obtained for the
mass drop tagger (and the modified mass-drop (mMDT) ) as well as for
pruning. It reflects that at this order the action of Y-splitter, in
the small $\rho$ limit, is to remove a logarithm in $\rho$ and replace
it with a (smaller) logarithm in $y$. This implies a reduction in the
QCD background at small $\rho$ relative to the plain jet mass
result. For $\rho > y$, the cut is redundant and we return to the case
of the plain QCD jet mass.

It is also straightforward to extend the soft approximation by
considering hard-collinear corrections. To include these effects one
simply makes the replacement $\frac{1}{x} \to \frac{1+(1-x)^2}{2x}$
i.e. includes the full QCD $p_{gq}$ splitting function. It is also
simple to include finite $y$ corrections in the above result by
inserting the proper limits of integration that are obtained from the
Y-splitter condition when one considers hard collinear rather than
soft gluon emission. The Y-splitter condition is satisfied for $y/(1-y)
< x <1/(1+y)$ and we obtain the result, for $\rho<y/(1+y)$:
\begin{equation}
\label{eq:finycut}
\frac{\rho}{\sigma} \frac{d\sigma}{d\rho}^{\mathrm{LO},\mathrm{coll.}} 
 = \frac{C_F \alpha_s}{\pi} 
   \left(\ln\frac{1}{y} - \frac{3}{4} \left( \frac{1-y}{1+y} \right)\right).
\end{equation}
This result is again identical to the case of (m)MDT with the $\ycut>y$
condition \cite{Dasgupta:2013via}.

\subsection{NLO result and all-orders form}\label{sec:pure-nlo}

Here we shall compute the next-to-leading order result in the soft and
collinear limit, before extending this result to all orders in the
next section.

Thus we need to consider the case of two real emissions off the
primary hard parton as well as a real emission and a virtual gluon
also treated in the soft and collinear limit. We shall work in the
classical independent emission approximation which is sufficient to
obtain the leading logarithmic result for jet mass distributions.

We consider a jet made up of a primary hard parton and two soft gluons
with four-momenta $k_1$ and $k_2$. When the jet is declustered one
requires the Y-splitter cut to be satisfied for the jet to be
tagged. There are two distinct situations that arise at this order:
firstly the situation where the largest $k_t$ gluon passes the
Y-splitter cut as well as sets the mass of the jet and secondly where
the largest $k_t$ gluon passes the Y-splitter cut so the jet is
accepted but the jet mass is set by a lower $k_t$ emission.

For the one-real, one-virtual contributions the
situation is the same as that for the leading order calculation
i.e. the real emission both passes the Y-splitter cut and sets the
mass.

 Let us assume that the jet mass is set by emission $k_1$ with
energy fraction $x_1$ and which makes an angle $\theta_1$ with the jet
axis or equivalently the hard parton direction, with $x_1,  \theta_1
\ll 1$.
For simplicity, it is useful to introduce for every emission $k_i$,
the quantities
\begin{equation}
\kappa_i \equiv x_i\theta_i,\qquad
\rho_i \equiv x_i\theta_i^2,
\end{equation}
respectively related to the transverse momentum ($k_t$ scale) of
emission $k_i$ wrt the jet axis and the contribution of emission $k_i$
to the jet mass.
We can then write 
\begin{align}
\label{eq:constraints}
\frac{1}{\sigma}\frac{d\sigma}{d \rho}^{\mathrm{NLO, soft-coll.}}=&
\left(\frac{C_F \alpha_s}{\pi} \right)^2 \int  d\Phi_2\,
   \delta \left ( \rho-\rho_1 \right)
   \bigg[ \Theta \left(\kappa_1>\kappa_2 \right)
           \Theta \left(x_1>y \right) 
           \Theta \left(\rho_2<\rho \right)  \nonumber \\ 
  & + \Theta \left(\kappa_2>\kappa_1 \right)
      \Theta \left(\kappa_2>\rho y \right)
      \Theta \left(\rho_2<\rho\right) 
    - \Theta \left(x_1 >y \right)  \bigg],
\end{align}
where we introduced the notation
\begin{equation}
d\Phi_2 \equiv \frac{dx_1}{x_1} \frac{dx_2}{x_2}
               \frac{d\theta_1^2}{\theta_1^2}\frac{d\theta_2^2}{\theta_2^2},
\end{equation}
for the two-gluon emission phase space in the soft-collinear limit.

The first line within the large parenthesis expresses the condition
that the gluon which sets the mass has the higher $k_t$
i.e. $\kappa_1(\equiv x_1 \theta_1) > \kappa_2 (\equiv x_2 \theta_2)$
as well as satisfies the Y-splitter constraint on the higher $k_t$ gluon
$\kappa_1^2/\rho_1=x_1^2 \theta_1^2/(x_1 \theta_1^2) =x_1 > y$. The
emission $k_2$ cannot dominate the jet mass by assumption, which gives
rise to the veto condition $\rho_2 < \rho$.  The first term on
the second line within the parenthesis expresses the condition that
the gluon $k_1$ now has lower $k_t$ than emission $k_2$. Emission
$k_2$ passes the Y-splitter cut $\kappa_2^2/\rho>y$, where
$\rho$ is the mass set by emission $k_1$. The final term on the last
line, with negative sign, is the contribution where emission $k_2$ is
virtual. 

For the term on the first line we make the replacement
$\Theta \left(\kappa_1>\kappa_2 \right) =1-\Theta \left(\kappa_2
  >\kappa_1 \right)$.
These two terms can be combined with the virtual corrections and the
first term of the second line, respectively, to give
\begin{multline}
\label{eq:split}
\frac{1}{\sigma}\frac{d\sigma}{d \rho}^{\mathrm{NLO, soft-coll.}} =
\left(\frac{C_F \alpha_s}{\pi} \right)^2 
  \bigg[ \int d\Phi_2\, \delta\left(\rho_1-\rho\right)
                        \Theta\left(x_1>y \right)
          \left(\Theta \left(\rho_2<\rho \right)-1 \right) \\
+ \int  d\Phi_2 \, \delta\left(\rho-\rho_1 \right)
   \Theta \left(\kappa_2>\kappa_1 \right) 
     \Theta \left(\rho_2<\rho \right)
   \left[\Theta \left(\kappa_2>y\rho \right)
       - \Theta \left(x_1>y \right) \right] \bigg].
\end{multline}

The fundamental reason for writing the result in the above form is to
separate what we expect to be the leading logarithmic contribution in
the first line from subleading contributions which involve a higher
$k_t$ emission giving a smaller contribution to the jet mass than
emission $k_1$. Hence we anticipate that the term in the second line
in Eq.~\eqref{eq:split} will produce results that are beyond our
accuracy, in the limit of small $\rho$. On explicit calculation of
this term one gets, for $\rho < y$,
\begin{align}
\label{eq:yssub}
\left (\frac{C_F \alpha_s}{\pi} \right)^2 &\int  d \Phi_2 \, 
   \delta\left(\rho_1-\rho \right)
   \Theta \left(\kappa_2>\kappa_1 \right) 
   \Theta \left(\rho_2<\rho \right)
   \left[\Theta \left(\kappa_2>y\rho \right)
       - \Theta \left(x_1>y \right) \right] \nonumber \\
&= \left (\frac{C_F \alpha_s}{\pi} \right)^2 \frac{1}{2\rho} 
   \left(\ln\frac{1}{\rho} \ln^2 \frac{1}{y}- \ln^3 \frac{1}{y} \right)
 = \left (\frac{C_F \alpha_s}{\pi} \right)^2 \frac{1}{2\rho} 
   \,\ln\frac{y}{\rho} \,\ln^2 \frac{1}{y}.
\end{align}

The above result implies that in the $\rho \to 0$ limit there are at
best single logarithmic (in $\rho$) contributions to the integrated jet mass
distribution from the second line of Eq.~\eqref{eq:split}.
Using $\Theta(\rho_2<\rho)-1=-\Theta(\rho_2>\rho)$, the first line of
Eq.~\eqref{eq:split} gives
\begin{equation}
\label{eq:twoemissions}
\frac{1}{\sigma}\frac{d\sigma}{d \rho}^{\mathrm{NLO,LL}} 
  = - \left(\frac{C_F \alpha_s}{\pi}\right)^2 
    \int  d\Phi_2 \Theta \left(x_1>y \right)
                  \delta \left ( \rho-\rho_1 \right)
                  \Theta \left(\rho_2 >\rho\right),
\end{equation}
which produces the leading logarithmic (LL)  corrections
we require. Upon evaluation, it produces for $\rho < y$,
\begin{equation}
\label{eq:nloll}
 \frac{\rho}{\sigma}\frac{d\sigma}{d \rho}^{\mathrm{NLO,LL}}=-\left ( \frac{C_F \alpha_s}{\pi} \right)^2 \frac{1}{2} \ln
 \frac{1}{y} \ln^2 \frac{1}{\rho},
\end{equation}
which has the structure of the leading-order result multiplied by a
double logarithmic term in $\rho$. 
We note that for $\rho > y$ the Y-splitter cut becomes redundant and one
returns to the result for the standard plain jet mass distribution.
We recall that by ``leading logarithmic (LL) accuracy'' we mean that
we only keep the terms that are maximally enhanced in $\ln\rho$. 

The result in Eq.~\eqref{eq:nloll} has a simple physical
interpretation. The largest $k_t$ emission which sets the mass comes
with a cut on its energy precisely as at leading order which, produces
an $\alpha_s \ln \frac{1}{y}$ behaviour. Emission $k_2$ on the other
hand is subject to a veto condition such that $\rho_2 < \rho$. After
cancellation against virtual corrections one obtains an
$\alpha_s \ln^2 \frac{1}{\rho}$ behaviour from this emission, exactly
as for the leading order contribution to the integrated plain jet mass
distribution. Based on this we can expect that at all orders, to
leading-logarithmic accuracy, one ought to multiply the leading-order
(LO) result by a double logarithmic Sudakov suppression factor like
that for the plain jet mass. The leading order result then appears as
a single-logarithmic prefactor in front of a resummed
double-logarithmic Sudakov exponent, as we shall see in the next
section.

Lastly we note that the full result of our calculation of
Eq.~\eqref{eq:constraints} can be written in the form

\begin{equation}
\label{eq:llplly}
\frac{1}{\sigma} \frac{d\sigma}{d\rho}^{\mathrm{NLO,soft-coll.}} =
\left ( \frac{C_F \alpha_s}{\pi} \right)^2 \frac{1}{2\rho}
\left(-\ln\frac{1}{y} \ln^2 \frac{1}{\rho} + \ln \frac{y}{\rho}
  \ln^2 \frac{1}{y}\right),
\end{equation}
where the first term on the RHS contains the leading logarithms in
$\rho$ while the second term is subleading in $\rho$ (being purely
single logarithmic), although it is enhanced by logarithmic terms in $y$.

\subsection{All-orders resummation and comparison to Monte Carlo results}\label{sec:pure-resum}

\begin{figure}
  \centerline{%
    \includegraphics[width=0.4\textwidth]{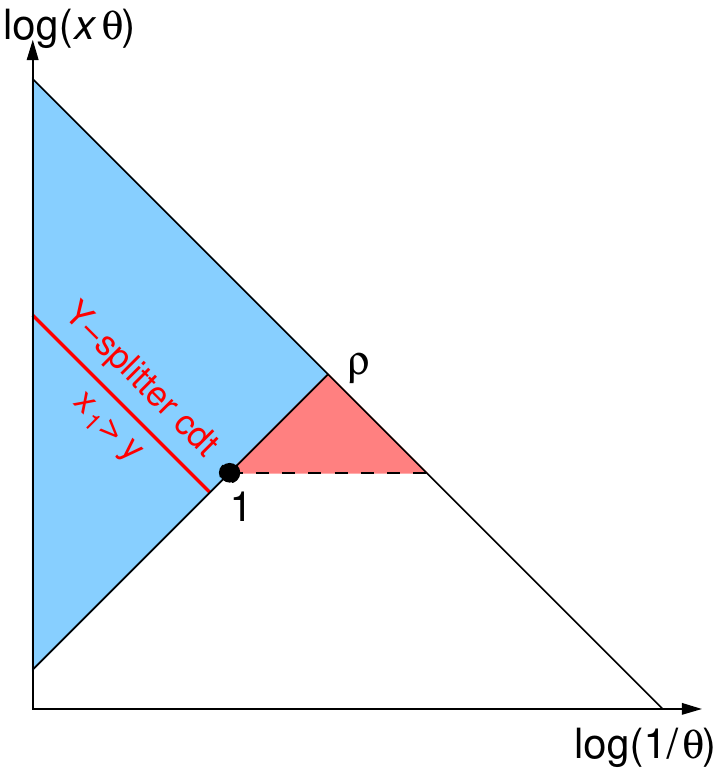}%
    \hfill%
    \includegraphics[width=0.4\textwidth]{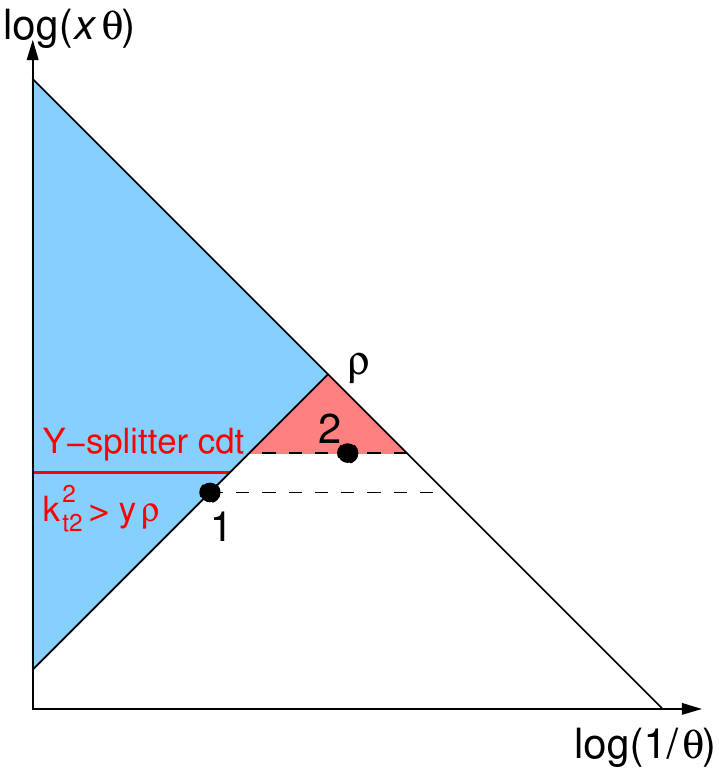}%
  }
  \caption{Lund diagrams representing the two contributions to the
    all-ordered resummed mass distribution. Left: the emission that
    dominates the jet mass also has the largest $k_t$; right: there
    is an emission with larger $k_t$ than the $k_t$ of the emissions
    which dominates the mass.}\label{fig:lund-pure}
\end{figure}

Eqs.~\eqref{eq:twoemissions}, \eqref{eq:nloll} can be easily
generalised to all orders.
To LL accuracy, one has to consider only the situation
where the highest $k_t$ emission dominates the jet mass. A jet-mass
veto then applies to all other real emissions.
This situation is depicted in the figure (``Lund diagram'') to the
left in Fig.~\ref{fig:lund-pure}. 
The emission denoted with a black dot sets the jet mass i.e. satisfies
$\rho_1\equiv x_1 \theta_1^2 =\rho$. The blue shaded region
corresponds to emissions that give a contribution to the mass
$x \theta^2 > \rho$ and hence are vetoed.  Considering these emissions
to be emitted according to an ``independent emission'' pattern the
veto condition gives a Sudakov suppression factor represented by the
blue shaded area in the figure which is identical to the suppression
factor obtained for the plain jet mass at leading-logarithmic
accuracy. In addition to this, emissions with a higher transverse
momentum which set a lower mass than $\rho$ are also vetoed since we
assumed that the emission which sets the mass is the highest $k_t$
emission. This is denoted by the red shaded area in the figure but as
this region produces only terms that are subleading in $\rho$ we shall
not consider it for the moment. Finally, we also have to consider the
Y-splitter constraint which for this configuration corresponds to
$x_1 > y$ where the line $x=y$ is shown in red in the figure.  The
all-orders fixed-coupling result from this configuration, which
captures the leading double-logarithms in $\rho$, is
\begin{equation}\label{eq:YSall}
\frac{\rho}{\sigma} \frac{d\sigma}{d\rho}^{\text{LL}}
   = \frac{C_F \alpha_s}{\pi} \ln \frac{1}{y} \times
     \exp \left [-\frac{C_F \alpha_s}{2\pi} \ln^2 \frac{1}{\rho} \right ],
     \quad (\text{for }\rho < y),
\end{equation}
while for $\rho > y$ the result is that for the plain mass
distribution.  Eq.~\eqref{eq:YSall} corresponds to the result reported
already in Eq.~\eqref{eq:basicform} and quoted in Ref.~\cite{Dasgupta:2015yua}. 
Note that a similar result is obtained also for the case of Y-pruning in the
regime $\alpha_s \ln \frac{1}{z_{\mathrm{cut}}} \ln \frac{1}{\rho}
\ll1$ (see Eq.~5.10b of Ref.~\cite{Dasgupta:2013ihk}).

It is simple to include running-coupling corrections both in the
prefactor i.e. those associated to the emission which sets the mass as
well as in the Sudakov exponent. Likewise hard-collinear emissions may
be treated by using the full splitting function in the prefactor and
the Sudakov exponent, yielding the modified leading logarithmic
approximation. Lastly we can also include finite $y$ corrections into
the prefactor as they may be of numerical significance since they
occur already at leading order (see Eq.~\eqref{eq:finycut}).

The general result, for $\rho<y$  then reads\footnote{Note that
  here and henceforth we shall only specify the transition points in a small
  $y$ approximation. Thus the exact transition point
  $\rho=y/(1+y)$ will be approximated by $\rho =y$.}

\begin{equation}\label{eq:pure-resum-leading}
\frac{\rho}{\sigma} \frac{d\sigma}{d\rho}^{\text{LL}} = 
\int_{\frac{y}{1+y}}^{\frac{1}{1+y}}dx_1\,P(x_1)\,\frac{\alpha_s(x_1\rho)}{2\pi}
e^{-R_{\text{plain}}(\rho)} ,
\end{equation}
where we defined the Sudakov exponent (``radiator'')
\begin{equation}\label{eq:radiator-rho}
R_{\text{plain}}(\rho) = \int \frac{d\theta^2}{\theta^2}dx\,P(x)\,
  \frac{\alpha_s(x^2 \theta^2)}{2\pi}
  \Theta \left(x \theta^2>\rho \right).
\end{equation}
and one has $P(x_1) = C_F p_{gq}(x_1)$ for quark jets, while identical
considerations hold for gluon jets with use of the appropriate
splitting functions for gluon branching to gluons and quarks.
In the above expression and the remainder of the text, the arguments
of the running coupling have to be understood as factors of
$p_t^2R^2$.
Explicit expressions for $R_{\text{plain}}$ as well as for all the other Sudakov
exponents used for the analytic results and plots in this paper are
given in Appendix~\ref{app:radiators}. 

In the present case, if $y$ becomes small enough, we can also perform
an all-order resummation of the logarithms of $1/y$. Such terms, which
are formally at the level of subleading logarithms in $\rho$, were
already identified in our fixed-order NLO calculation, see
Eq.~\eqref{eq:llplly}.  In order to resum them we will have to
consider also situations where the highest transverse momentum
emission does not set the jet mass. To write a general resummed result
it is convenient to return to the Lund diagrams in
Fig.~\ref{fig:lund-pure}. The figure on the left denotes, as we
stated before, the situation where the highest transverse momentum
emission both passes the Y-splitter constraint and also sets the mass,
with a veto on higher mass emissions. Now however we also account for
the contribution from the red shaded region that corresponds to an
additional veto on emissions with a higher transverse momentum than
the emission which sets the mass. The figure on the right denotes a
second situation where there is an emission $k_2$ which is the highest
$k_t$ emission i.e. $\kappa_2 >\kappa_1$. The red shaded region now
denotes the additional veto on any emissions with transverse momentum
greater than $\kappa_2$. The blue region as before corresponds to a
veto on emissions with larger mass than $\rho=\rho_1$ and the
Y-splitter condition now corresponds to $\kappa_2^2 > \rho y$
where the line $x^2 \theta^2 =\rho y$ is shown in the figure.
\begin{figure}
  \centerline{%
    \includegraphics[width=0.49\textwidth]{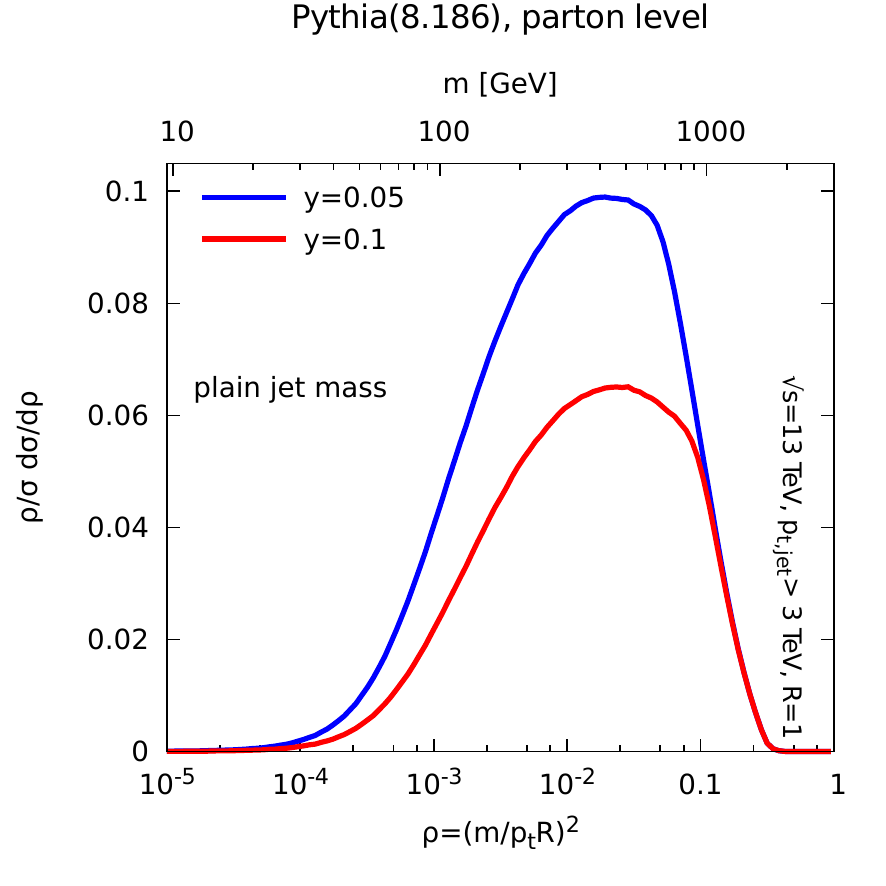}%
    \hfill
    \includegraphics[width=0.49\textwidth]{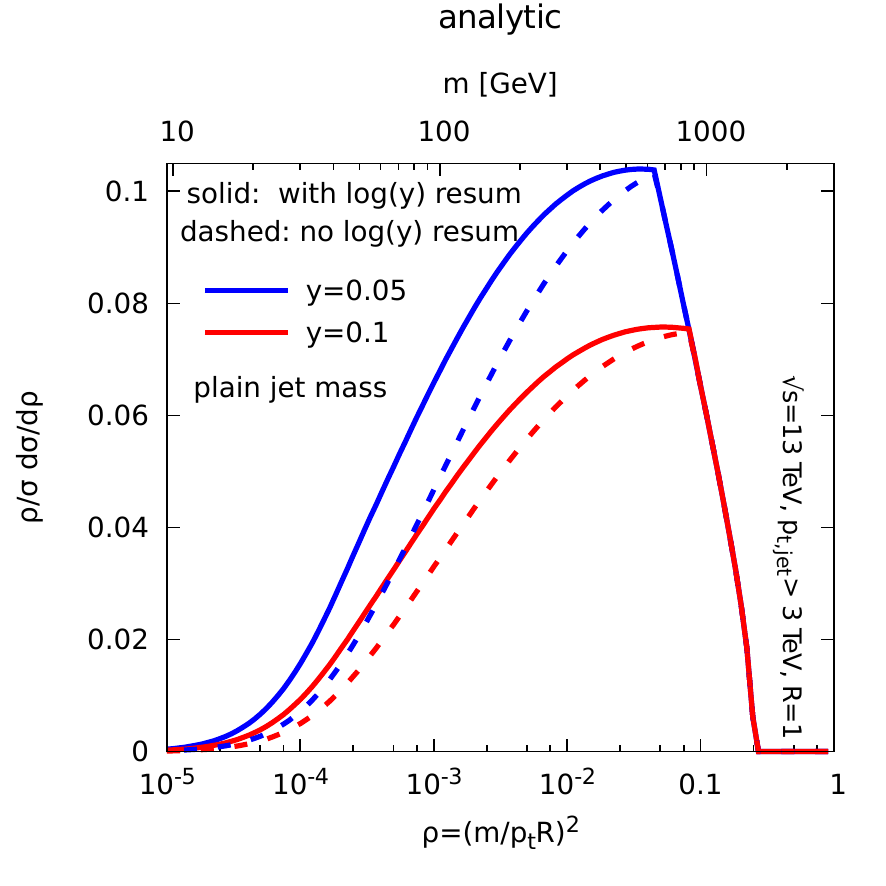}%
  }
  \caption{Comparison of our analytic results (right) with Pythia
    simulations (left) for two values of $y$. For the analytic
    curves, the solid lines correspond to
    Eq.~(\ref{eq:pure-basic-resum}), \ie include the all-order
    resummation of the logarithms of $y$, while the dashed lines
    correspond to Eq.~(\ref{eq:pure-resum-leading}), \ie do not
    include the resummation of the logarithms of $y$
  }\label{fig:pure-mcplot}
\end{figure}

Taking both the above described situations into account one
can write the result as (for now we ignore finite $y$ effects to
which we shall return)
\begin{align}\label{eq:pure-basic-resum}
\frac{\rho}{\sigma} \frac{d\sigma}{d\rho}^{\text{LL+LL}_y} & =
  \int_\rho^1 dx_1\, P(x_1) \,\frac{\alpha_s(\rho x_1)}{2\pi} 
  e^{-R_{\text{plain}}(\rho)}
  \Bigg[
     \Theta (x_1>y)
     e^{-R_{k_t}(\kappa_1,\,\rho)} \, + \\ \nonumber
  &+ \, 
  \int \frac{d\theta_2^2}{\theta_2^2}dx_2\,P(x_2)\,
  \frac{\alpha_s(\kappa_2^2)}{2\pi} 
  \Theta\left(\rho_2<\rho\right)
  \Theta\left(\kappa_2 > \rho x_1\right)
  \Theta\left(\kappa_2 > \rho y \right)
  e^{-R_{k_t}(\kappa_2,\,\rho)} \Bigg ] ,
\end{align}
where the first term in large brackets comes from the Lund diagram on
the left and the second term from that on the right. 
Note that $R_{k_t}$ is also a Sudakov type exponent defined as
\begin{equation}\label{eq:radiator-kt}
R_{k_t}(\kappa,\rho) = \int \frac{d\theta^2}{\theta^2}dx\,P(x)\,
  \frac{\alpha_s(x^2 \theta^2)}{2\pi}
  \Theta \left(x \theta^2<\rho \right)
  \Theta \left(x \theta>\kappa \right),
\end{equation}
which arises from a veto on transverse momentum of emissions above the
scale $k_t$ while at the same time imposing that the mass of the
vetoed emissions is lower than $\rho$, as required for taking into
account the red shaded regions in the Lund diagrams of Fig.~\ref{fig:lund-pure}.

This expression can be simplified quite significantly: one first splits
the second line into a contribution with $x_1>y$ and a contribution
with $\rho<x_1<y$. After integration over $x_2$ and $\theta_2$ and
combining the contribution from $x_1>y$ with the first line of
(\ref{eq:pure-basic-resum}) one can write the final result as 
\begin{equation}\label{eq:pure-resum-all-logs}
\frac{\rho}{\sigma} \frac{d\sigma}{d\rho}^{\text{LL+LL}_y}
  = e^{-R_{\text{plain}}(\rho)}\,\left[
    \int_{\frac{y}{1+y}}^{\frac{1}{1+y}}dx_1\,P(x_1)
    \frac{\alpha_s(x_1\rho)}{2\pi}
  +\left(1-e^{-R_{k_t}(\sqrt{\rho y},\rho)}\right)
   \int_\rho^y dx_1\,P(x_1)\frac{\alpha_s(x_1\rho)}{2\pi}\right]\,,
\end{equation}
where we have restored the finite $y$ corrections in the leading
contribution (first term).
The correction term one thus obtains relative to
(\ref{eq:pure-resum-leading}) has a prefactor proportional to
$\alpha_s\ln \frac{y}{\rho}$ multiplied by a Sudakov-like factor,
starting at order $\alpha_s$ and resumming terms of the form
$\alpha_s^n\ln^{2n}\frac{1}{y}$. This is consistent with the result
obtained at NLO in Eq.~\eqref{eq:llplly}.

In order to validate our analytic results, we have compared them to
Monte-Carlo simulations. We have used Pythia (v8.186)~\cite{pythia}
with the 4C tune~\cite{Corke:2010yf} to generate $qq\to qq$ events at
parton level with $\sqrt{s}=13$~TeV.
Jets are reconstructed with the anti-$k_t$ algorithm~\cite{antikt}
with $R=1$ as implemented in
FastJet~\cite{Cacciari:2005hq,Cacciari:2011ma} and we require that the
jets satisfy $p_t>3$~TeV and rapidity $|y|<4$.
Unless explicitly mentioned otherwise, the same setup is used for all
the subsequent Monte-Carlo simulations in this paper.

The comparison to our analytic calculations is shown in Figure
\ref{fig:pure-mcplot} with Pythia on the left and our results on the
right.
All our results include the contribution from the full splitting
function including hard-collinear effects to the Sudakov exponent, and
use a 1-loop approximation for the running of the strong coupling with
$\alpha_s(M_Z)=0.1383$. This value matches the one used in Pythia for
the final-state shower.
Furthermore, the plot with our analytic results includes
both the leading logarithmic result described in
Eq.~\eqref{eq:pure-resum-leading} (dashed curves) as well as the
result augmented to include resummation of double logarithms in $y$,
Eq.~\eqref{eq:pure-basic-resum} (solid curves) for two values of
$y$. We note firstly the good overall agreement with Monte Carlo
results for both variants of the analytics, which indicates that our
modified leading-logarithmic results successfully explain the
performance of Y-splitter on QCD background jets. The observed
differences between analytics and Monte Carlo can arise due to
different treatment of next-to--leading logarithmic effects such as
those due to soft emissions at large angles and initial state
radiation included in the Monte Carlo studies but left out of our
resummed calculations.

It is noteworthy that the $\ln y$ resummation although a visible effect,
is fairly modest. The essential dependence of the results on $y$ is
already captured by the leading-logarithmic resummation of
Eq.~\eqref{eq:pure-resum-leading}.

\section{Y-splitter with grooming}\label{sec:grooming}

In this section we shall consider the Y-splitter method supplemented
with grooming procedures, specifically the modified mass-drop tagger
(equivalently SoftDrop $\beta=0$) and trimming. The effectiveness of
applying grooming subsequent to the use of Y-splitter on a jet has
been clearly demonstrated in the Monte Carlo studies carried out in
Ref.~\cite{Dasgupta:2015yua}. There it was shown that while Y-splitter
alone has a very poor signal efficiency (similar to that for an
ungroomed jet which is severely affected by ISR and underlying event),
grooming makes a considerable difference to the performance of
Y-splitter on signal jets. On the other hand we have already seen that
on QCD background jets Y-splitter gives a double-logarithmic Sudakov
type factor multiplying a single logarithmic prefactor, which implies
a desirable strong suppression of background. As already mentioned in
the introduction, the key observation made
in Ref.~\cite{Dasgupta:2015yua} was that using Y-splitter with
grooming did not significantly alter the performance of Y-splitter on
background jets, in the sense that applying a grooming procedure after
one imposes a $Y$-splitter cut does not alter the
double-logarithmic Sudakov behaviour for the QCD background.
This fact coupled with the great improvement seen in signal efficiency
resulted in Y-splitter+grooming outperforming other standard taggers
for signal significance at high $p_t$ . Here we seek to understand
from a first principles viewpoint why grooming does not appear to
strongly impact the basic performance of Y-splitter on background.  We
start by studying Y-splitter with trimming in the next sub-section, which
was the combination employed in Ref.~\cite{Dasgupta:2015yua}.

\subsection{Y-splitter with trimming: fixed-order results}\label{sec:trimming-fo}

To study the impact of trimming on Y-splitter, we shall consider
taking a jet accepted by Y-splitter and then apply trimming to it.
It is important to highlight that it is crucial to apply the
Y-splitter condition on the plain jet and apply grooming
afterwards. We show in Appendix~\ref{app:grooming+Ysplitter} that
applying grooming first and then imposing the Y-splitter condition on
the groomed jet leads to a smaller suppression of the QCD background.

We shall set the $f_{\mathrm{cut}}$ parameter of trimming to be equal
to the parameter $y$ of Y-splitter, a choice that will become clear
presently.
\footnote{If we keep into account finite $y$ corrections,
  we should actually use $f_{\rm cut}=y/(1+y)$, which is what we have
  done in practice in our Monte Carlo simulations.} We firstly note
that, at leading order, for a soft emission to pass Y-splitter it must
have an energy fraction $x>y$. When one applies trimming afterwards
such an emission is unaffected as, with our choice of
$f_{\mathrm{cut}}$ trimming removes only emissions with
$x<y$. Thus at leading-order Y-splitter with trimming trivially
returns the same result as Y-splitter alone.

We shall now examine the role of trimming at the NLO level.
Let us consider that the mass of the final jet \emph{after} grooming
is set by an emission $k_1$. In other words, we first impose the
Y-splitter cut on the plain jet and, if it passes, we compute the
trimmed jet mass.

At order $\alpha_s^2$ we have to consider both a second real emission
$k_2$ as well as a virtual gluon contribution.
The mass distribution can be written as\footnote{Since we explained
  the approximations we have made in the previous section we shall no
  longer explicitly specify that the NLO corrections here are computed
  in the limit of soft and collinear emissions.}
\begin{equation}
\frac{1}{\sigma}\frac{d\sigma}{d\rho}^{\mathrm{NLO,soft-coll}} = \left (\frac{C_F \alpha_s}{\pi} \right)^2
\int d\Phi_2
\left(\mathcal{I}_1+\mathcal{I}_2+\mathcal{I}_3+\mathcal{I}_4 \right)
\end{equation}
with
\begin{align}
\mathcal{I}_1 
  &= \delta\left(\rho-\rho_1\right)
     \Theta\left(\kappa_1>\kappa_2 \right)
     \Theta\left(\frac{\kappa_1^2}{\rho_1+\rho_2}>y\right) 
     \Theta\left(\rho_2<\rho \right) 
     \Theta_2^{\mathrm{in}},\label{eq:integ1}\\
\mathcal{I}_2
  &= \delta\left(\rho-\rho_1\right)  
     \Theta\left(\kappa_1>\kappa_2\right)
     \Theta\left(\frac{\kappa_1^2}{\rho_1+\rho_2}>y\right)
     \Theta_2^{\mathrm{out}}, \label{eq:integ2}\\
\mathcal{I}_3
  &= \delta\left(\rho-\rho_1\right)
     \Theta\left(\kappa_2>\kappa_1\right)
     \Theta\left(\frac{\kappa_2^2}{\rho_1+\rho_2}>y\right) 
     \Theta\left(\rho_2<\rho\right) 
     \Theta_1^{\mathrm{in}},  \label{eq:integ3}\\
\mathcal{I}_4
  &=- \delta \left(\rho-\rho_1\right) \Theta(x_1>y),
\end{align}
where we introduced the shorthand notations $\Theta_i^{\mathrm{in}}$
and $\Theta_i^{\mathrm{out}}$ to represent that emission $k_i$ is
respectively left in or removed by trimming. We recall the condition
for an emission to be removed by trimming is
\begin{equation}
\Theta_i^{\mathrm{out}} 
  = 1 - \Theta_i^{\mathrm{in}}
  = \Theta(x_i<y)\,\Theta(\theta_i>r),
\end{equation}
with $r \equiv \frac{R_{\mathrm{trim}}}{R}$ and $R_{\mathrm{trim}}$
the trimming radius.

Let us detail the physical origin of these different contributions.
The contribution $\mathcal{I}_1$ contains the conditions on
$x_1,x_2,\theta_1,\theta_2$ such that $k_1$ sets the mass
($\rho= \rho_1$) and has the higher transverse momentum, $\kappa_1 >
\kappa_2$. 
It also contains the condition for the
Y-splitter cut to pass $\kappa_1^2/(\rho_1+\rho_2)>y$, and the
condition that $k_2$ is left in by trimming represented by
$\Theta_2 ^{\mathrm{in}}$. Lastly it contains the veto on the mass
$\rho > \rho_2$ such that emission $k_2$ cannot set the mass.
Likewise $\mathcal{I}_2$ contains the conditions that emerge when
$k_2$ is removed by trimming which itself corresponds to the condition
$\Theta_2^{\mathrm{out}}$.
For both $\mathcal{I}_1$ and $\mathcal{I}_2$, the Y-splitter condition
implies $x_1>y$ and therefore guarantees that emission $k_1$ is left
in by trimming.
These configurations reproduce the leading-logarithmic terms of the
pure Y-splitter cut, and also generate subleading contributions coming
from the region where $k_2$ is removed by trimming and has
$\rho_2>\rho$.\footnote{One can easily see this by inserting
  $1=\Theta(\rho_2>\rho)+\Theta(\rho_2<\rho)$ in $\mathcal{I}_2$.}
$\mathcal{I}_3$ represents the situation when $k_1$ is the lower
transverse momentum emission and sets the mass. In this case, the
Y-splitter condition implies $x_2>y$, \ie emission $k_2$ is kept by
trimming, and we thus have to impose that $\rho_2<\rho_1$. We also
have to impose that emission $k_1$ is left in by trimming
corresponding to $\Theta_1^{\mathrm{in}}$.
Lastly $\mathcal{I}_4$ corresponds to the situation when $k_2$ is
virtual and all that is required is for $k_1$ to pass the Y-splitter
cut.

A comment is due about the Y-splitter condition used in the above
formulae Eqs.~\eqref{eq:integ1} --\eqref{eq:integ3}. In situations
where emission $k_1$ dominates the mass even though emission $k_2$ is
not groomed away it is possible, at leading logarithmic accuracy, to
replace $\rho_1+\rho_2$ in the denominator of the Y-splitter
constraints by $\rho=\rho_1$.  Specifically this applies to the 
$\mathcal{I}_1$ and $\mathcal{I}_3$ terms above. We have however
chosen to treat the Y-splitter constraint exactly in all terms since
in the term involving $\mathcal{I}_2$, where emission $k_2$ is groomed
away, there is no condition on $\rho_2$ requiring it to be less
than $\rho$.  Retaining the exact Y-splitter constraint in all terms
proves convenient for reorganising and combining various contributions as
we shall do below, while only differing from the leading-logarithmic
simplification by subleading terms which we do not control.

Given that one of the main observations motivating this work is that
the use of grooming techniques does not drastically modify the
background rejection obtained with Y-splitter alone, it is of interest
to express the calculations as grooming-induced corrections to those
already carried out for Y-splitter. To this end, in the contribution
involving $\mathcal{I}_1$ let us replace $\Theta_2^{\mathrm{in}}$ with
$1-\Theta_2^{\mathrm{out}}$ which splits the contribution from
$\mathcal{I}_1$ into two pieces $\mathcal{I}_1=
\mathcal{I}_1^{\mathrm{full}} -\mathcal{I}_1^{\mathrm{out}}$.  The
contribution from $\mathcal{I}_1^{\mathrm{full}}$, where we can use
$\rho_1+\rho_2\approx \rho_1$ in the Y-splitter condition, is just the
same as the corresponding leading term for the pure Y-splitter
case. It can be combined with the virtual term $\mathcal{I}_4$ (which
is also identical to the pure Y-splitter case) to produce the NLO
leading-logarithmic result we reported earlier for Y-splitter,
cf. Eqs.~(\ref{eq:YSall}) and ~(\ref{eq:pure-resum-leading}). We can
apply a similar procedure for the term $\mathcal{I}_3$ such that
$\mathcal{I}_3= \mathcal{I}_3^{\mathrm{full}}
-\mathcal{I}_3^{\mathrm{out}}$, where $\mathcal{I}_3^{\mathrm{full}}$
is the contribution to the pure Y-splitter case from the situation
that the the highest $k_t$ emission passes Y-splitter but does not set
the jet mass. Recall that this configuration produces only terms
beyond our formal leading-logarithmic accuracy (cf. the second term in
Eq.~(\ref{eq:pure-resum-all-logs})).  The remaining terms, all
involving $\Theta_2^{\mathrm{out}}$, constitute the trimming-induced
corrections to Y-splitter.
It is then useful to write the result in the following form:
\begin{equation}
\frac{1}{\sigma}  \frac{d\sigma}{d\rho}^{\mathrm{NLO,soft-coll}} =
\frac{1}{\sigma}\frac{d\sigma}{d\rho}^{{\mathrm{NLO,YS}}}+\mathcal{F}^{\mathrm{trim,a}} +\mathcal{F}^{\mathrm{trim,b}}
\end{equation}
where $\frac{1}{\sigma}\frac{d\sigma}{d\rho}^{{\mathrm{NLO,YS}}}$ is
the pure Y-splitter result given by
Eq.~(\ref{eq:pure-resum-all-logs}), and we defined
\begin{equation}
\label{eq:ftrima}
\mathcal{F}^{\mathrm{trim,a}}
  = \left(\frac{C_F\alpha_s}{\pi}\right)^2
    \int d\Phi_2 \delta\left(\rho-\rho_1\right)
    \Theta\left(\kappa_1>\kappa_2\right)
    \Theta\left(\frac{\kappa_1^2}{\rho_1+\rho_2}>y\right)
    \left [1-\Theta\left(\rho_2<\rho \right) \right]
    \Theta_2^{\mathrm{out}},
\end{equation}
which arises from combining the contributions from
$\mathcal{I}_2$  and $-\mathcal{I}_1^{\mathrm{out}}$ and 
\begin{equation}\label{eq:ftrimb}
\mathcal{F}^{\mathrm{trim,b}}
  = -\left(\frac{C_F\alpha_s}{\pi}\right)^2
    \int d\Phi_2 \delta\left(\rho-\rho_1\right)
   \Theta\left(\kappa_2>\kappa_1\right)
   \Theta\left(\frac{\kappa_2^2}{\rho_1+\rho_2}>y\right)
   \Theta\left(\rho_2<\rho\right) \Theta_1^{\mathrm{out}},
\end{equation}
which arises from the $-\mathcal{I}_3^{\mathrm{out}}$ term. 

At this stage, within our accuracy we can replace $\rho_1+\rho_2$ by
$\rho_2$ in (\ref{eq:ftrima}) and by $\rho_1$ in (\ref{eq:ftrimb}).
We can then express the constraints in Eq.~\eqref{eq:ftrima} in the form
\begin{equation}
\label{eq:k2out}
\delta \left(\rho-\rho_1\right)
\Theta\left(\frac{\rho x_1}{x_2}>\rho_2 \right)
\Theta\left(\frac{\rho x_1}{y}>\rho_2 \right)  
\left [1-\Theta \left(\rho_2<\rho \right) \right]
\Theta_2^{\mathrm{out}}.
\end{equation}

We note that the above implies the condition $x_1 >y$ and
$\Theta_2^{\mathrm{out}}$ imposes the condition $x_2 < y$ since
emission $k_2$ has to be removed by trimming. Thus we have that
$x_1/x_2 >x_1/y$.
As a consequence Eq.~\eqref{eq:k2out} can be written as
\begin{equation}\label{eq:k2out2}
\delta \left(\rho-\rho_1\right)\left[  \Theta\left(\rho_2<\frac{\rho x_1}{y}\right)
 -\Theta\left(\rho_2<\rho\right)
  \Theta\left(\rho_2<\frac{\rho x_1}{y} \right) \right]\Theta_2^{\mathrm{out}}.
\end{equation}

For $x_1 < y$ this vanishes while for $x_1 > y$ the term in big square
brackets gives
$\Theta \big(\rho_2<\frac{\rho x_1}{y}\big)  -\Theta
\left(\rho_2<\rho \right) $.
Thus one finally gets for $\mathcal{F}^{\mathrm{trim,a}}$
\begin{equation}
\label{eq:YScorr}
\mathcal{F}^{\mathrm{trim,a}}
 = \left(\frac{C_F\alpha_s}{\pi} \right)^2 
   \int d\Phi_2  \,\Theta_2^{\mathrm{out}} 
   \delta\left(\rho-\rho_1\right) 
   \Theta\left(x_1>y \right)
   \left[ \Theta\left(\rho_2<\frac{\rho x_1}{y}\right)
         -\Theta\left(\rho_2<\rho\right) \right].
\end{equation}

The above result has a simple interpretation. The veto on emissions
that one places for the case of pure Y-splitter is modified by the
action of trimming.  In the region where emissions are removed by
trimming, emissions are no longer subject to the direct constraint that the
mass must be less than $\rho$, which represents the subtraction of the
$\Theta \left(\rho_2 < \rho \right)$ veto condition in the $\Theta_2^{\mathrm{out}}$ region. However emissions in this region, even
though they are removed by trimming, are still subject to the
constraint $k_{t1}^2/m_j^2 >y$ which is the Y-splitter cut
and where $m_j^2$ is the squared invariant mass of the ungroomed jet, to which
all emissions, including those removed eventually by grooming, do
contribute. Thus one gets the correction to pure Y-splitter given by
Eq.~\eqref{eq:YScorr}, from those configurations where the highest
$k_t$ emission sets the final jet mass. \footnote{These, we recall, are
  the configurations that generate the leading logarithmic corrections
  for pure Y-splitter.} 

It is simple to calculate $\mathcal{F}^{\mathrm{trim,a(b)}}$. The form of
the result depends on the value of $\rho$ and there are various
regimes that emerge. In what follows we shall choose values such that
$r^2 < y$, as is common for phenomenological purposes, although our
main conclusions will be unchanged by making a different choice.  One has:
\begin{itemize}
\item \underline{The regime $\rho < y^2 r^2$}\nopagebreak

Here we find
\begin{align}
\label{eq:smallrho}
&\mathcal{F}^{\mathrm{trim,a}} 
  = \frac{1}{\rho}\left (\frac{C_F \alpha_s}{\pi}\right)^2 
    \frac{1}{2} \ln \frac{1}{r^2}\ln^2 y\\
\label{eq:ftrimbsmallrho}
&\mathcal{F}^{\mathrm{trim,b}} 
  = -\frac{1}{\rho}\left (\frac{C_F \alpha_s}{\pi}\right)^2
    \frac{1}{2} \ln\frac{1}{r^2} \ln^2 y \\
&\mathcal{F}^{\mathrm{trim,a}}+\mathcal{F}^{\mathrm{trim,b}} = 0.
\label{eq:ftrimab0}
\end{align}

The above results are noteworthy since they indicate that in the small
$\rho$ limit, $\rho \to 0$, where one may regard resummation of
logarithms of $\rho$ to be most important, the overall correction to
Y-splitter vanishes at our leading-logarithmic accuracy. This is also
the essential reason for the fact that trimming does not appear to
significantly modify the performance of Y-splitter on background jets,
as the basic structure of a Sudakov form factor suppression at small
$\rho$ is left unchanged.

\item \underline{The regime $ y^2 r^2 < \rho < y r^2$}\nopagebreak

One obtains
\begin{equation}
\label{eq:ftrimaint}
\mathcal{F}^{\mathrm{trim,a}}
 = \frac{1}{\rho}\left(\frac{C_F \alpha_s}{\pi}\right)^2 \bigg(
   \frac{1}{2} \ln^2\frac{1}{y}\ln\frac{1}{r^2}
 - \frac{1}{6} \ln^3\frac{\rho}{y^2r^2}\bigg),
 \end{equation}
while for $\mathcal{F}^{\mathrm{trim,b}}$ the result coincides with
that quoted in Eq.~ \eqref{eq:ftrimbsmallrho}. Thus we have for the
full correction from trimming:

\begin{equation}
\label{eq:ftrimab1}
\mathcal{F}^{\mathrm{trim,a}}+\mathcal{F}^{\mathrm{trim,b}} =
-\frac{1}{\rho}\left(\frac{C_F \alpha_s}{\pi} \right)^2 \frac{1}{6}
\ln^3 \frac{\rho}{y^2 r^2}.
\end{equation}

It is instructive to examine the behaviour of Eq.~\eqref{eq:ftrimab1} at the
transition points: for $\rho = y^2 r^2$ it vanishes and hence trivially matches onto
Eq.~\eqref{eq:ftrimab0} while for $\rho = yr^2$ we get
\begin{equation}
\label{eq:rhointermediate1}
-\frac{1}{\rho} \left(\frac{C_F \alpha_s}{\pi}\right)^2 \frac{1}{6} \ln^3 \frac{1}{y}.
\end{equation}

\item \underline{The regime $ y^2 > \rho > y r^2 $}\nopagebreak

Here one gets
\begin{equation}
\mathcal{F}^{\mathrm{trim,a}}
 = \frac{1}{\rho}\left(\frac{C_F \alpha_s}{\pi} \right)^2 \left(
   \frac{1}{2} \ln\frac{y}{\rho} \ln^2 \frac{1}{y}
 - \frac{1}{6} \ln^3\frac{1}{y} \right).
\end{equation}

On the other hand the result for $\mathcal{F}^{\mathrm{trim,b}} $ in this
region is 
\begin{equation}
\label{eq:ftrimblargep}
\mathcal{F}^{\mathrm{trim,b}}
  =-\frac{1}{\rho}\left(\frac{C_F \alpha_s}{\pi} \right)^2 
  \frac{1}{2} \ln\frac{y}{\rho} \ln^2 \frac{1}{y},
\end{equation}
such that 
\begin{equation}
\label{eq:rhoindep}
\mathcal{F}^{\mathrm{trim,a}} +\mathcal{F}^{\mathrm{trim,b}}
  = - \frac{1}{\rho}\left(\frac{C_F \alpha_s}{\pi} \right)^2
    \frac{1}{6} \ln^3 \frac{1}{y},
\end{equation}
i.e. independent of $\rho$.

Note that the above result is identical to that reported in
Eq.~\eqref{eq:rhointermediate1} for $\rho = y r^2$ as one would expect.

\item \underline{The regime $y>\rho>y^2$}\nopagebreak

Here one obtains
\begin{equation}
\label{eq:trimintermediate}
\mathcal{F}^{\mathrm{trim,a}}
 = \frac{1}{\rho} \left(\frac{C_F \alpha_s}{\pi} \right)^2 \left(
   \frac{1}{3} \ln^3\frac{y}{\rho}
 + \frac{1}{2} \ln^2\frac{y}{\rho} \ln\frac{\rho}{y^2} \right).
\end{equation}
The result for $\mathcal{F}^{\mathrm{trim,b}}$ in this region remains
the same as in Eq.~\eqref{eq:ftrimblargep} so that 
\begin{equation}
\label{eq:faplusfbint}
\mathcal{F}^{\mathrm{trim,a}}+\mathcal{F}^{\mathrm{trim,b}} =
\frac{1}{\rho} \left(\frac{C_F \alpha_s}{\pi} \right)^2
\ln\frac{y}{\rho}
  \left(\frac{5}{6}\ln \frac{1}{\rho} \ln \frac{1}{y}
 -\frac{7}{6} \ln^2 \frac{1}{y}
 -\frac{1}{6} \ln^2 \frac{1}{\rho}\right),
\end{equation}
which matches on to Eq.~\eqref{eq:rhoindep} at $\rho=y^2$ and vanishes
at $\rho=y$.

\end{itemize}

For $\rho >y$ the functions $\mathcal{F}^{\mathrm{trim,a(b)}}$ vanish
and there is no correction to Y-splitter which itself coincides with
the plain jet mass.

To summarise, we find that, in the formal small $\rho$ limit, we recover the
same result as for the pure Y-splitter case at this order (see the
region $\rho<y^2r^2$).  As we move towards larger values of $\rho$
i.e. beyond  $\rho = y^2 r^2$, we find that the result becomes
substantially more complicated. We find transition points at $y^2 r^2$,
$y r^2$, $y^2$ and $y$ which arise due to the use of trimming. The
result in all these regions contains logarithms of $\rho$ along with
logarithms of $y$ ( as well as $\ln r$ terms) . However in these
regions logarithms of $\rho$ cannot be considered to be dominant over
other logarithms such as those in $y$.  To get a better feeling for
the size of the corrections to the pure Y-splitter case in various
regions it is helpful to look at the behaviour at the transition
points. At $\rho =y^2 r^2$ the correction due to trimming vanishes
while at $\rho =yr^2$ one finds an overall correction varying as
$\frac{1}{\rho} \alpha_s^2 \ln^3 y$ which is formally well beyond our
leading-logarithmic accuracy in $\rho$, although enhanced by logarithms
of $y$. The behaviour at other transition points is similarly highly
subleading in $\rho$ though containing logarithms in $y$.  As we have
already noted before resummation of $\ln y$ enhanced terms has only a
modest effect and does not affect our understanding of the basic
behaviour of the tagger (see Fig.~\ref{fig:pure-mcplot}). 

The fixed-order results of this section already explain why the action
of trimming following the application of Y-splitter only changes the
performance of Y-splitter at a subleading level. It is simple to carry out
a resummed calculation valid at the leading logarithmic level in
$\rho$ but with only an approximate treatment of subleading
terms. Such a resummed calculation is in fact seen to be in
qualitative agreement with Monte Carlo studies.  However a feature of
the result obtained with trimming, which is perhaps undesirable from
a phenomenological viewpoint, is the position of multiple transition
points in the final result. While these transition points are not as
visible as for the case of pure trimming itself (see
Ref.~\cite{Dasgupta:2013ihk}) it may nevertheless be desirable to
think of using grooming methods which are known to have less
transition points in conjunction with Y-splitter. To this end we shall
first investigate the modified mass drop tagger (mMDT)  at fixed-order
before addressing the question of resummation and comparisons to Monte
Carlo of Y-splitter with grooming.

\subsection{Y-splitter with mMDT: fixed-order results}\label{sec:mmdt}

The NLO calculation for Y-splitter with mMDT proceeds similarly to the
case of the Y-splitter trimming combination but with differences of
detail. If one considers the correction to the pure Y-splitter case at
this order, we arrive at functions $\mathcal{F}^{\mathrm{mMDT,a(b)}}$
which can be computed exactly like $\mathcal{F}^{\mathrm{trim,a(b)}}$
with the only difference being in the condition
$\Theta_2^{\mathrm{out}}$ for removal of emission $k_2$ by the mMDT as
well as condition $\Theta_1^{\mathrm{in}}=1-\Theta_1^{\mathrm{out}}$
which differs from the trimming case.  To be more explicit, for mMDT
to remove the emission $k_2$ one has that
$\Theta_2^{\mathrm{out}} =\Theta \left(\theta_2>\theta_1 \right)
\Theta \left(x_2<y \right)$
since mMDT would not reach emission $k_2$ if it were at smaller angle
than $k_1$, as $k_1$ passes the mMDT cut.

In contrast to trimming, the final result contains only two transition
points at for $\rho = y^2$ and $\rho=y$. We obtain for the correction
to Y-splitter 
  $\mathcal{F}^{\mathrm{mMDT}} =
  \mathcal{F}^{\mathrm{mMDT,a}}+\mathcal{F}^{\mathrm{mMDT,b}}$ such that:

\begin{itemize}
\item \underline{For $\rho<y^2$}\nopagebreak
\begin{equation}\label{eq:mmdt-small-rho-region}
\mathcal{F}^{\mathrm{mMDT}} = -\frac{1}{\rho} \left(\frac{C_F \alpha_s}{\pi} \right)^2\frac{1}{6} \ln^3 \frac{1}{y}.
\end{equation}
This agrees with the result for trimming at $y r^2<\rho<y^2$, quoted
in Eq.~\eqref{eq:rhoindep}.

\item \underline{For $y>\rho>y^2$}\nopagebreak

Here again the result is identical to that obtained for trimming
i.e. the sum of $\mathcal{F}^{\mathrm{trim,a}}$ and
$\mathcal{F}^{\mathrm{trim,b}}$ in the same region. 
\end{itemize}
Note that one can alternatively obtain the mMDT results by taking the
limit $r\to 0$ in the trimming results.

As before, for $\rho>y$ one obtains no correction from grooming or
Y-splitter and the result for the plain mass is recovered, meaning
once more that grooming will not substantially affect the small-$\rho$
behaviour of Y-splitter.

In summary using mMDT as a groomer produces a result that, as for the
case of trimming, produces only subleading corrections in terms of
logarithms of $\rho$ and hence leaves the pure Y-splitter Sudakov
unaltered at leading logarithmic level in the limit of small
$\rho$. The subleading terms carry enhancements involving logarithms
of $y$ as for trimming, but there are fewer transition points for mMDT
than trimming, which is certainly a desirable feature from a
phenomenological viewpoint.

\subsection{All-orders calculation and comparisons to Monte-Carlo results}\label{sec:mmdt-resum}

As explicitly shown via fixed-order calculations in the previous
section, the use of grooming methods subsequent to the application of
Y-splitter does not modify the leading logarithmic results in a small
$\rho$ resummation. It is straightforward  to see that this
statement extends beyond fixed-order to all perturbative orders and is
the reason why previous Monte Carlo studies \cite{Dasgupta:2015yua}
observed that the performance of Y-splitter on background jets is not fundamentally altered by groomers.

Beyond the leading logarithmic level however the situation with
Y-splitter becomes more complicated when one introduces grooming. For
trimming there are multiple transition points that are obtained in
addition to the transition point at $\rho=y$, which is already present
for pure Y-splitter. For values of $\rho$ which are larger than
$y^2 r^2$, the structure of the results is complicated and logarithms
of $\rho$ can no longer be considered dominant. One may therefore
wonder about the practical impact of such formally subleading
corrections on the tagger behaviour.  It is therefore of some interest
to write down a resummed result that goes beyond leading-logarithmic
accuracy in $\rho$ and captures some of the formally subleading terms
that emerge in the various regimes we have identified, such as those
enhanced by logarithms of $y$.

It proves to be relatively straightforward to carry out the same kind
of resummation as reflected by Eqs.~ (\ref{eq:pure-basic-resum}) and
(\ref{eq:pure-resum-all-logs}) for the pure Y-splitter case, which
retain both leading logarithms in $\rho$ and those in $y$.  In
Appendix \ref{sec:mMDTYSresum} we carry out a resummed calculation
along these lines for Y-splitter with mMDT. The result we obtain is:
\begin{align}\label{eq:mmdt-resum}
\frac{\sigma}{\rho}\frac{d\sigma}{d\rho}^{\text{LL+LL$_y$}}
 & = 
 \int_y^1 dx_1\,P(x_1)\frac{\alpha_s(\rho x_1)}{2\pi} 
   e^{-R_{\text{plain}}(\rho)}\\
 &\!\!\!\!\!\!\!\!\!
   \left[e^{-R_{k_t}(\kappa_1;\rho)-(R_{\text{out}}(\kappa_1^2/y)-R_{\text{out}}(\rho))}
   + \int_{\kappa_1}^{\sqrt{\rho}} \frac{d\kappa_2}{\kappa_2}
   R_{k_t}'(\kappa_2;\rho)e^{-R_{k_t}(\kappa_2;\rho)-(R_{\text{out}}(\kappa_2^2/y)-R_{\text{out}}(\rho))}\right],\nonumber
\end{align}
where $R_{\text{plain}}(\rho)$ and  $R_{k_t}(\kappa;\rho)$ are defined in
Eqs.~(\ref{eq:radiator-rho}) and (\ref{eq:radiator-kt}) respectively,
and 
\begin{equation}
 R_{\text{out}}(\rho)-R_{\text{out}}(\kappa_1^2/y) =
   \int\frac{d\theta^2}{\theta^2}dx\,P(x)\,\frac{\alpha_s(x^2\theta^2)}{2\pi}
   \Theta(x<y)\,\Theta(\kappa_1^2/y>x\theta^2>\rho).
\end{equation}
One can fairly easily show that the second line in
\eqref{eq:mmdt-resum} only brings subleading logarithmic contributions
(in $\ln\rho$), so that the LL result is fully given by the first line
in \eqref{eq:mmdt-resum} and corresponds to the LL result for pure
Y-splitter.
This can be obtained from the following observations.
The $R_{k_t}$ factors, already encountered before, bring at most
subleading corrections proportional to $\alpha_s\ln^2 y$. 
Then, since $\kappa_1^2/y=\rho x_1/y$ and $y<x_1<1$,
$R_{\text{out}}(\rho)-R_{\text{out}}(\kappa_1^2/y)$ can at most bring
single-logarithmic corrections proportional to $\alpha_s \ln\rho\,\ln y$.
This remains valid for
$R_{\text{out}}(\rho)-R_{\text{out}}(\kappa_2^2/y)$ since
$\ln(\kappa_1^2/\kappa_2^2)$ can at most introduce logarithms of $y$
(see Appendix \ref{sec:mMDTYSresum} for more details) .

Alternatively, it is instructive to evaluate (\ref{eq:mmdt-resum})
with a fixed-coupling approximation.
Assuming, for simplicity, that $\rho<y^2$, and working in the
soft-collinear approximation where we can use $P(x)=2C_F/x$, we have
\begin{align}
R_{k_t}'(\kappa_i;\rho) & =\frac{2\alpha_sC_F}{\pi}\ln\frac{\rho}{\kappa_i^2},\\
R_{k_t}(\kappa_i;\rho) & =\frac{\alpha_sC_F}{2\pi}\ln^2\frac{\rho}{\kappa_i^2},\\
R_{\text{out}}(\rho)-R_{\text{out}}(\kappa_i^2/y) & =\frac{\alpha_sC_F}{2\pi}\Big(\ln^2\frac{y}{\rho}-\ln^2\frac{y^2}{\kappa_i^2}\Big).
\end{align}
Substituting these expressions in Eq.~(\ref{eq:mmdt-resum}) one can
reach after a few manipulations 
\begin{align}\label{eq:resum-mmdt-fixed}
\frac{\sigma}{\rho}\frac{d\sigma}{d\rho}^{\text{LL+LL$_y$}}
  =  e^{-R_{\text{plain}}(\rho)}
 \int_y^1 \frac{dx}{x}\frac{\alpha_sC_F}{\pi}
  \bigg(1+\frac{\alpha_sC_F}{\pi}\ln\frac{1}{x}\ln\frac{x}{y}\bigg)
 e^{-\frac{\alpha_sC_F}{2\pi}\big(\ln^2x-\ln\frac{x}{y}\ln\frac{y^3}{\rho^2x}\big)}.
\end{align}
In the above expression, the factor in front of the exponential as
well as the first term in the exponential only yield terms of the form
$(\alpha_s\ln^2y)^n$, and the second term in the exponential will lead to
both  $(\alpha_s\ln^2y)^n$ and $(\alpha_s\ln y\ln\rho)^n$
contributions. These are both subleading compared to our desired
leading-logarithmic accuracy in $\rho$ so that
\eqref{eq:resum-mmdt-fixed} will lead to the
$\frac{\alpha_sC_F}{\pi}\ln\frac{1}{y}e^{-R_{\text{plain}(\rho)}}$ result plus subleading contributions as expected.

While a complete evaluation of the integral over $x$ in
\eqref{eq:resum-mmdt-fixed} is not particularly illuminating --- it
would give an error function --- it is interesting to expand it to
second order in $\alpha_s$. One obtains 
\begin{equation}
\frac{\sigma}{\rho}\frac{d\sigma}{d\rho}^{\text{LO+NLO,soft-coll}}
 = \frac{\alpha_sC_F}{\pi}\ln\frac{1}{y}
 -
 \frac{1}{2}\bigg(\frac{\alpha_sC_F}{\pi}\bigg)^2\ln\frac{1}{y}\bigg(\ln^2\rho-\ln\rho\ln
 y+\frac{4}{3}\ln^2y\bigg),
\end{equation}
which correctly reproduces the sum of (\ref{eq:llplly}) and
\~(\ref{eq:mmdt-small-rho-region}).

\begin{figure}
  \centerline{%
    \includegraphics[width=0.49\textwidth]{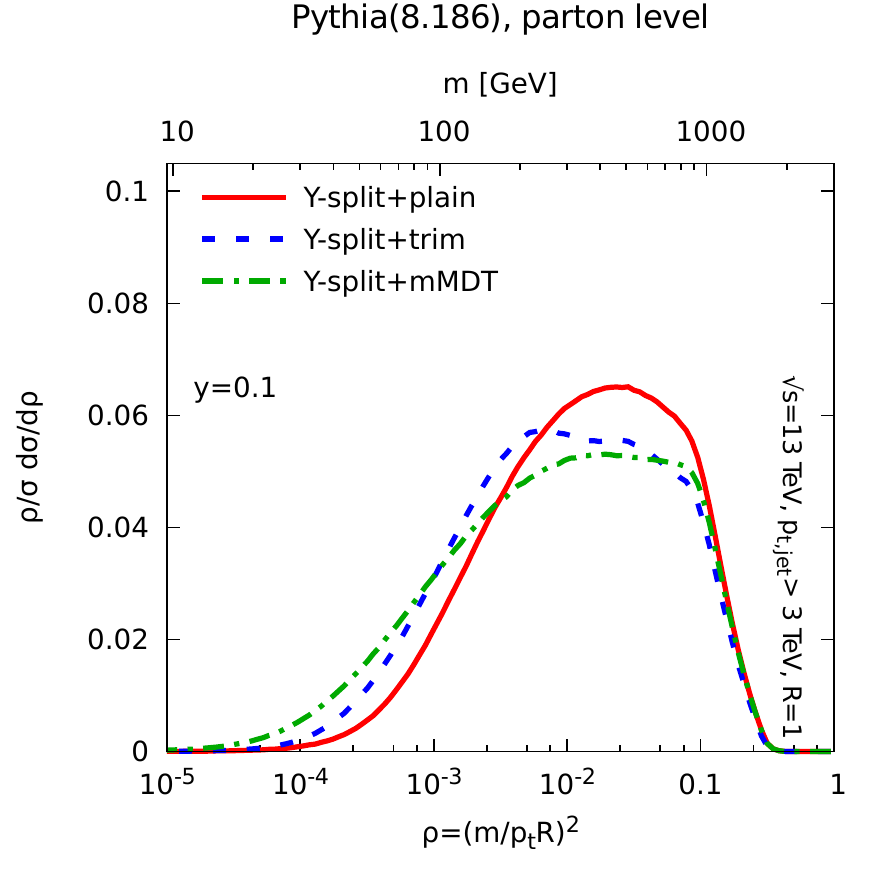}%
    \hfill
    \includegraphics[width=0.49\textwidth]{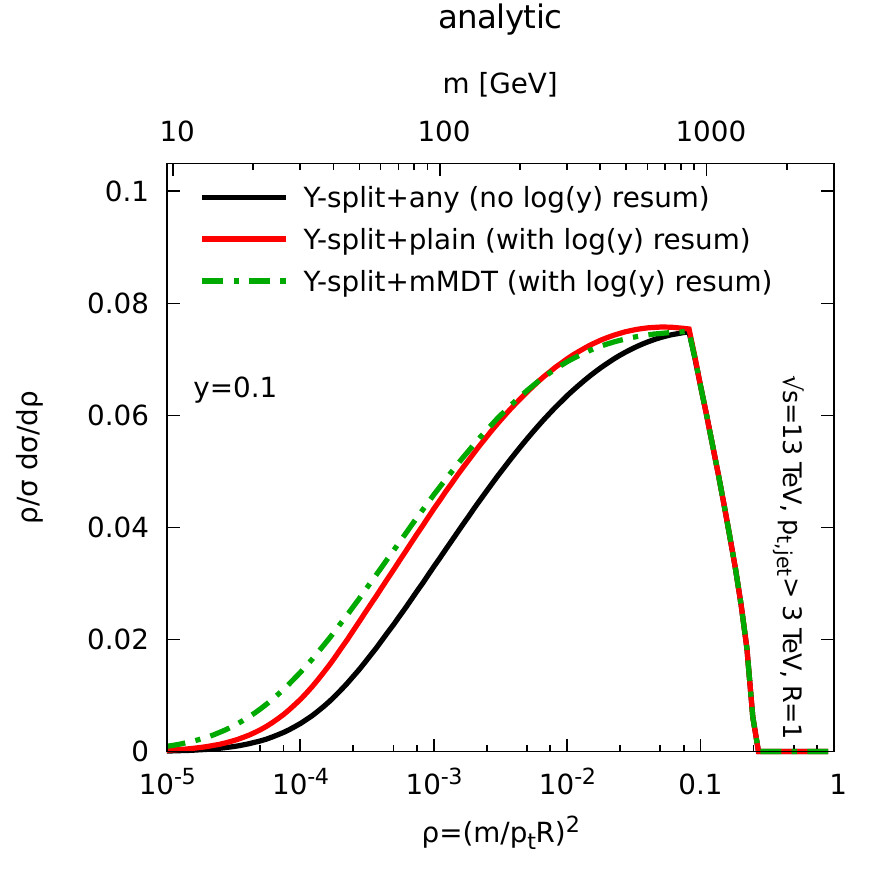}%
  }
  \caption{Comparison of our analytic results (right) with Pythia
    simulations (left) for different choices of grooming. For the
    analytic curves, we show the result including only the leading
    logarithms in $\rho$, Eq.~(\ref{eq:pure-resum-leading}),
    valid independently of the groomer, as well as the results
    including the resummation of the $\ln y$ terms for the
    pure Y-splitter case, Eq.~(\ref{eq:pure-resum-all-logs}), and the mMDT
    jet mass, Eq.~(\ref{eq:mmdt-resum}).}\label{fig:groomed-mcplot}
\end{figure}

Our result Eq.~\eqref{eq:mmdt-resum} shows that the leading
logarithmic results obtained for Y-splitter with mMDT coincide with
those for pure Y-splitter since the factor in the big square bracket
only generates subleading corrections to the pure Y-splitter
result. This result also contains the resummation of leading
logarithmic terms in $y$, which are subleading from the point of view
of $\ln \rho$ resummation. The analytic results for mMDT with $\ln y$
resummation are plotted in Fig.~\ref{fig:groomed-mcplot}. Also
plotted for reference is the leading logarithmic resummed result,
which is independent of whether we groom with mMDT or trimming, or not
at all. We can see that, as also observed before for the pure
Y-splitter case, resummation of $\ln y$ terms brings only modest
differences compared to the leading logarithmic answer. In
Fig.~\ref{fig:groomed-mcplot} the plot on the left shows the results
obtained with Monte Carlo studies for Y-splitter with trimming and
mMDT compared to pure Y-splitter.\footnote{We used the implementation
  of mMDT (and SoftDrop) provided in {\tt{fjcontrib}}
  \cite{fjcontrib}.} The plot reaffirms our observation that grooming
does not alter the essential feature of a Sudakov suppression at small
$\rho$. The Monte Carlo result for trimming also shows some hints of
the transition in behaviour induced by subleading terms and is
correspondingly less smooth than the mMDT result which has fewer
transition points.

We note that while we have performed a $\ln y$ resummation in order to
assess their impact on the LL result we do not claim that these terms
are numerically more important (for practically used values of $y$)
than other subleading in $\rho$ effects we have neglected, such as
non-global logarithms and multiple emission effects. Non-global
logarithms in particular are known to have a substantial impact on the
peak height of the jet-mass spectrum \cite{DassalNG1}. However these other
effects are harder to treat and hence we used the $\ln y$ resummation
as a convenient method to assess the impact of some subleading terms on the LL
result. 

\section{Variants}\label{sec:variants}

\subsection{Y-splitter with mass declustering}\label{sec:mass-ordering}

We have seen in the previous section that beyond the strict leading
logarithmic approximation in $\ln \frac{1}{\rho}$, the behaviour of the
tools can be quite complex, especially when we combine Y-splitter
with grooming. 
In this section, we discuss a small modification to the definition of
Y-splitter that largely simplifies this calculation and has the fringe
benefit of coming with a small performance enhancement.

Most of the complication in the calculations we have done so far comes
from the fact that the emission which  passes the
Y-splitter cut is the highest $k_t$ emission, which can be different
from the emission that dominates the mass. Such configurations produce
only terms beyond leading-logarithmic (LL)  accuracy but as we have seen
their structure is rather involved. The discussion and results beyond
LL would clearly be simpler if the $k_t$ scale entering Y-splitter was
directly calculated based on the emission that dominates the jet mass.
One can readily achieve this by {\em replacing the $k_t$ declustering by a
  generalised-$k_t$ declustering with $p=1/2$} which respects the
ordering in mass so that the emission that passes Y-splitter is also
the emission that dominates the jet mass.\footnote{A similar argument was already used
  in~\cite{Dasgupta:2015lxh} to compute the axes for
  $N$-subjettiness.} If we consider a soft emission with momentum fraction $x_1$ at
an angle $\theta_1$, which dominates the mass, this would give a cut of the form 
\begin{equation}\label{eq:Ym-split-basic}
\frac{x_1^2\theta_1^2}{x_1\theta_1^2} = x_1 >y.
\end{equation}
More precisely if we choose to include finite $y$ corrections one obtains
\begin{equation}\label{eq:Ym-split-basic+finite}
 \frac{\left(\mathrm{min} \left (x_1,1-x_1 \right)\right)^2\theta_1^2}{x_1(1-x_1)\theta_1^2} > y
\qquad \Rightarrow \qquad
\frac{1}{1+y}>x_1 > \frac{y}{1+y}.
\end{equation}

We denote this variant \Ym-splitter, where the subscript $\text{m}$
refers to the fact that we now use a mass-ordered declustering procedure. 
Regardless of whether we ultimately measure the jet mass without
grooming or the groomed jet mass, \Ym-splitter computed on the plain
jet will always impose that the emission that dominates the plain jet
mass has a momentum fraction larger than $y$.  In the case where we
measure the plain jet mass, we would therefore simply recover the
result quoted in (\ref{eq:pure-resum-leading}) with no
$\alpha_s^2\ln\frac{y}{\rho}\ln^2\frac{1}{y}$ correction.

On top of that, the \Ym-splitter condition guarantees that the emission
dominating the plain mass also passes the trimming (or mMDT) condition.
We would therefore also recover (\ref{eq:pure-resum-leading}) for the
\Ym-splitter+grooming case, as only emissions that do not essentially
affect the jet mass can be removed by grooming.

\begin{figure}
  \centerline{%
    \includegraphics[width=0.49\textwidth]{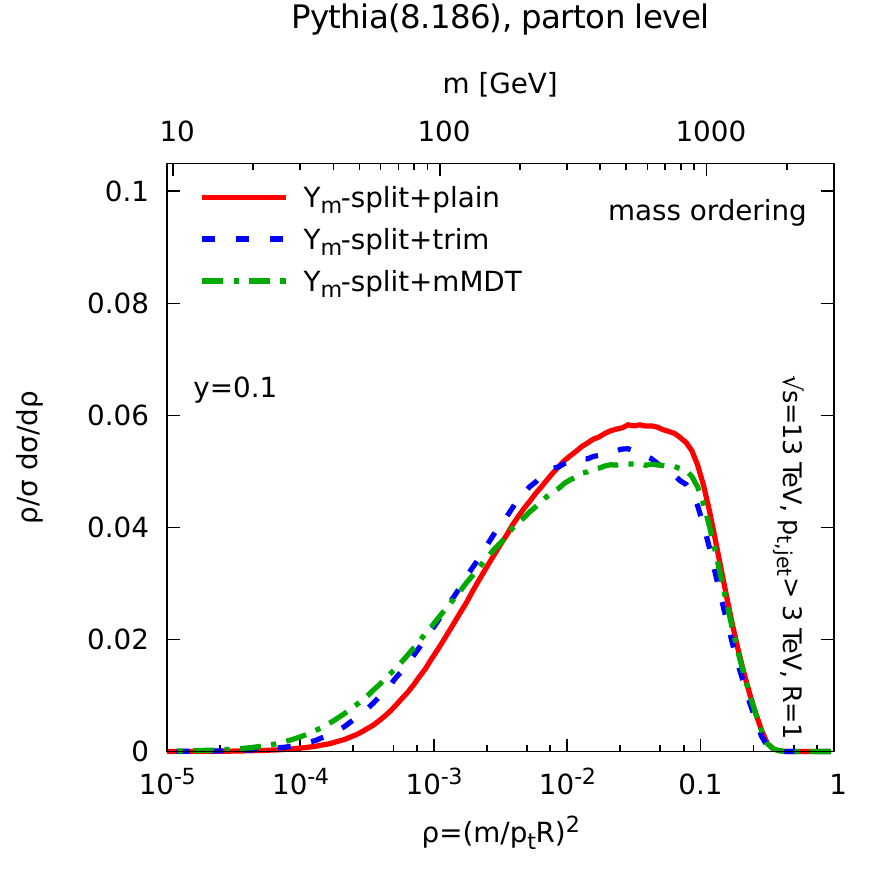}%
    \hfill
    \includegraphics[width=0.49\textwidth]{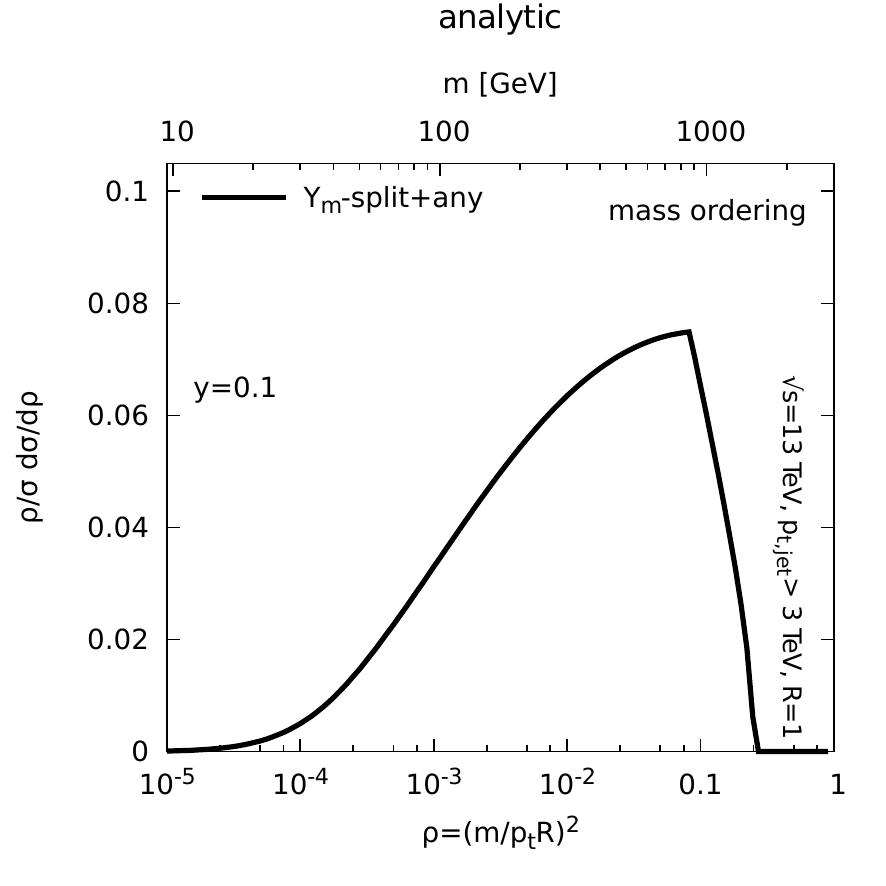}%
}
  \caption{Comparison of our analytic result
    Eq.~\eqref{eq:pure-resum-leading} (right) with Pythia simulations
    (left) for different choices of grooming for
    \Ym-splitter.}
\label{fig:mass-ordered-mcplot} 
\end{figure}

\begin{figure}
  \centerline{%
    \includegraphics[width=0.49\textwidth]{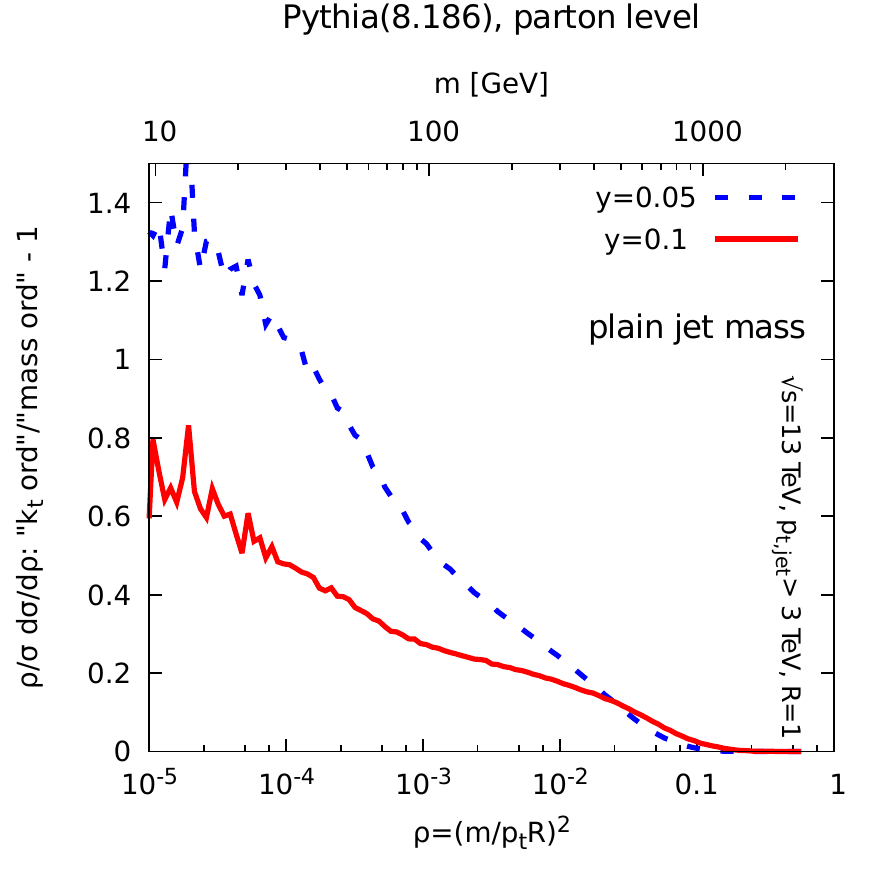}%
    \hfill
    \includegraphics[width=0.49\textwidth]{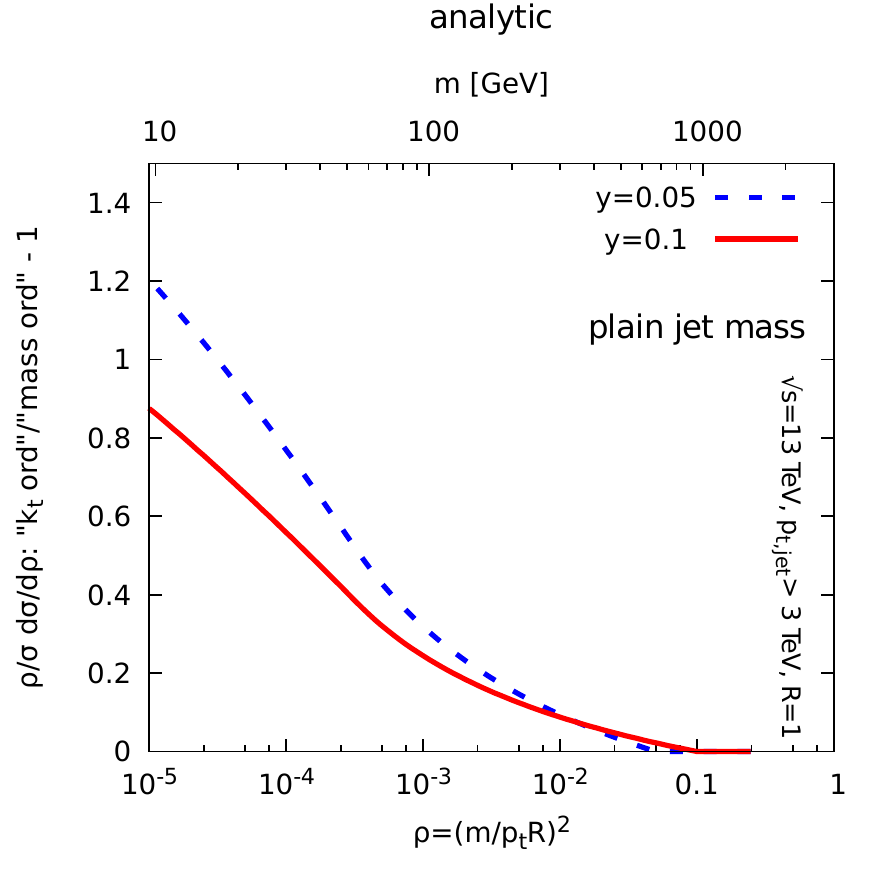}%
}
\caption{Ratio of mass distribution obtained with ($k_t$-ordered)
  Y-splitter divided by the mass distribution obtained with
  (mass-ordered) \Ym-splitter. We compare our analytic results (right)
  with Pythia simulations (left).}
  \label{fig:mass-v-kt-mcplot}
\end{figure}

Comparisons between Monte-Carlo simulations, still using Pythia8 at
parton level, and the analytic expectation
(\ref{eq:pure-resum-leading}) are presented in
Fig.~\ref{fig:mass-ordered-mcplot}. 
We clearly see that our analytic result captures very well the shape
observed in the Monte-Carlo simulation. It also appears that
differences between the ungroomed case and the two groomed cases are
smaller than what was observed for the standard Y-splitter case
discussed in the previous two sections (see \eg
Fig.~\ref{fig:groomed-mcplot}), as one would expect from the
analytical viewpoint. It appears also that using \Ym-splitter comes with a fringe benefit,
namely the fact that it suppresses the mass spectrum somewhat more than
Y-splitter does.
As an additional test of our analytic calculations, we can compare the
difference between our results for the mass-ordered case
Eq.~(\ref{eq:pure-resum-all-logs}) and
Eq.~(\ref{eq:pure-resum-leading}) representing our result for the
usual $k_t$ ordered Y-splitter to Monte-Carlo results. 
This is shown in Fig.~\ref{fig:mass-v-kt-mcplot} and, bearing in
mind that our analytic calculation only resums contributions maximally
enhanced by $\ln\frac{1}{y}$, shows a good agreement between the two sides
of the figure. 
Fig.~\ref{fig:mass-v-kt-mcplot} also illustrates the fact
that the difference between Y- and \Ym-splitter essentially behaves
like $\ln\frac{y}{\rho}$ up to running coupling corrections.

A comment is due about differences between the groomed and
ungroomed jet mass after imposing the \Ym-splitter condition. We would
still expect these differences to appear at subleading logarithmic orders in $\rho$
but they would not be enhanced by double logarithms of $y$.
It is also interesting to notice that while most of the NLL
corrections to the overall $\exp[-R_{\text{plain}}(\rho)]$ Sudakov
factor would be the same as for the plain jet mass, the correction due
to multiple emissions would be different.
This can be understood from the fact that, if several emissions,
$(x_1,\theta_1),\dots(x_n,\theta_n)$ contribute significantly to the
plain jet mass,  only the largest, say $(x_1,\theta_1)$, will be used
to compute the $k_t$ scale leading to the \Ym-splitter constraint
\begin{equation}\label{eq:Ym-split-multiple-emissions}
  x_1^2\theta_1^2>y\sum_{i=1}^nx_i\theta_i^2,
\end{equation}
which is no longer as simple as (\ref{eq:Ym-split-basic}), albeit more
constraining.
One can still carry out a resummation with this exact condition but it
leads to more complicated expressions which go beyond the scope of
this paper and beyond the accuracy we have aimed for here.
Note that at the same, single-logarithmic, order of accuracy, one
would anyway have to include additional contributions, in particular
the non-trivial contribution from non-global logarithms.

\subsection{Y-splitter with mass declustering and a $z$ cut}\label{sec:zcut}

It is possible to further simplify the analytic computations by having
the Y-splitter condition behave like a $\zcut$ rather than a $\ycut$,
in a spirit similar to what was proposed for the MassDropTagger in
\cite{Dasgupta:2013ihk}.\footnote{In the case of a $\zcut$-based
  \Ym-splitter, the mMDT and trimming would also use directly the
  parameter $z$ of \Ym-splitter as a momentum fraction cut.}
As before, we first decluster the jet using the generalised $k_t$
algorithm with $p=1/2$ to obtain two subjets $j_1$ and $j_2$. We then
impose the condition
\begin{equation}\label{eq:ysplitter-cdt-zcut}
\zcut \equiv \frac{\text{min}(p_{t1},p_{t2})}{p_{t1}+p_{t2}} > z.
\end{equation}

As for the case of a mass declustering with a $\ycut$, this would lead
to (\ref{eq:pure-resum-leading}) at leading logarithmic accuracy in
$\ln\frac{1}{\rho}$, and be free of subleading corrections enhanced by
logarithms of $z$.
Moreover, if multiple emissions, $(x_1,\theta_1),\dots(x_n,\theta_n)$,
contribute to the plain jet mass, with $x_1\theta_1^2\ge x_i\theta_i^2$,
the \Ym-splitter condition will give
\begin{equation}
\label{eq:Ym-zcut-multiple-emissions}
\zcut = x_1 > z.
\end{equation}
which is significantly simpler than the corresponding condition with a
$\ycut$, Eq.~(\ref{eq:Ym-split-multiple-emissions}).
This is valid independently of which mass, groomed or ungroomed, we
decide to measure.
However, even if we apply a grooming procedure, the \Ym-splitter
condition \eqref{eq:Ym-zcut-multiple-emissions} guarantees that the
emission $(x_1,\theta_1)$ which dominates the jet mass is kept by
grooming and dominates also the groomed jet mass.
The multiple-emission correction to the measured jet mass, groomed or
ungroomed, will therefore be sensitive to all the emissions, including
$(x_1,\theta_1)$, kept in the jet used to measure the mass. Their
resummation leads to the standard form \cite{Catani:1992ua} for
additive observables
$\exp(-\gamma_E R_{\text{mass}}')/\Gamma(1+R_{\text{mass}}')$, where
$R_{\text{mass}}'$ is the $\ln\frac{1}{\rho}$-derivative of the
Sudakov associated with the mass we consider \ie either the plain jet
mass or the groomed jet mass Sudakov.
The mass distribution is then given by
\begin{equation}\label{eq:ysplitter-zcut-with-me}
\frac{\rho}{\sigma} \frac{d\sigma}{d\rho}^{\text{LL+ME}} = 
\int_z^{1-z} dx_1\,P(x_1)\,\frac{\alpha_s(x_1\rho)}{2\pi} 
  \frac{e^{-R_{\text{plain}}(\rho)-\gamma_E
      R_{\text{mass}}'(\rho)}}{\Gamma \left(1+R_{\text{mass}}'(\rho) \right)},
\end{equation}
with the superscript ``ME'' indicating that the contribution from
multiple emissions is included and
\begin{equation}\label{eq:ysplitter-zcut-Rprime-mass}
R_{\text{mass}}'(\rho) = 
\int_0^1 \frac{d\theta^2}{\theta^2} dx\,P(x)\,\frac{\alpha_s(x^2\theta^2)}{2\pi} 
 \,\rho\delta(x\theta^2-\rho)\,\Theta_{\text{in}},
\end{equation}
where the $\Theta_{\text{in}}$ imposes that the emission is kept by
grooming, or is set to 1 for the plain jet mass.

\begin{figure}
  \centerline{%
    \includegraphics[width=0.49\textwidth]{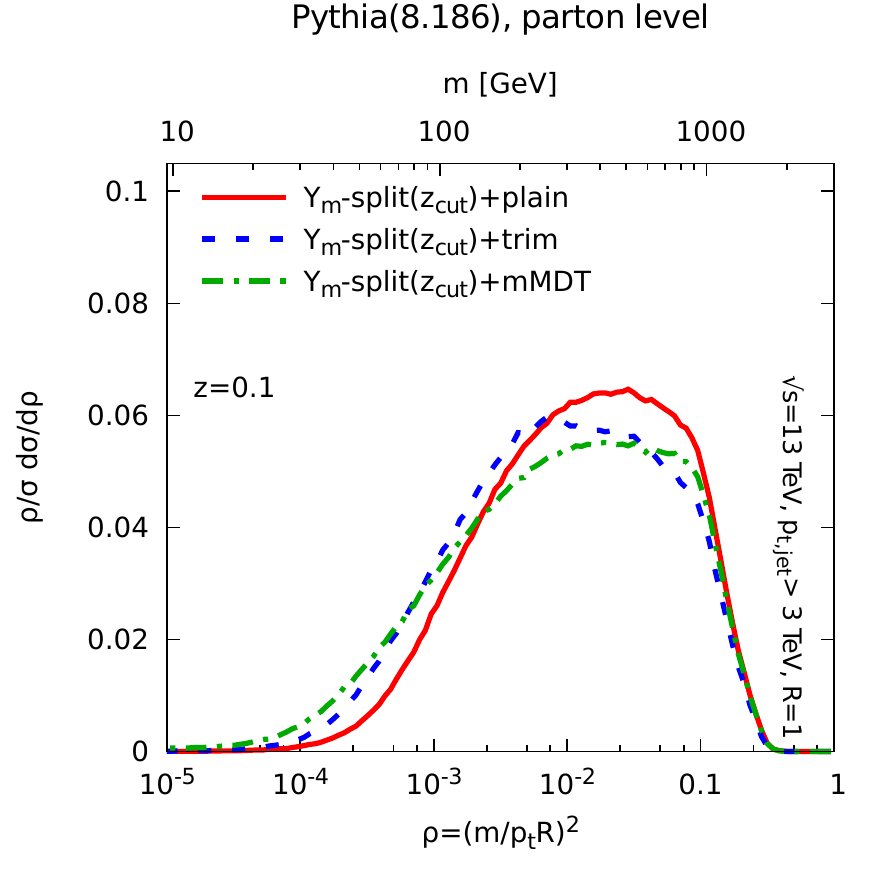}%
    \hfill
    \includegraphics[width=0.49\textwidth]{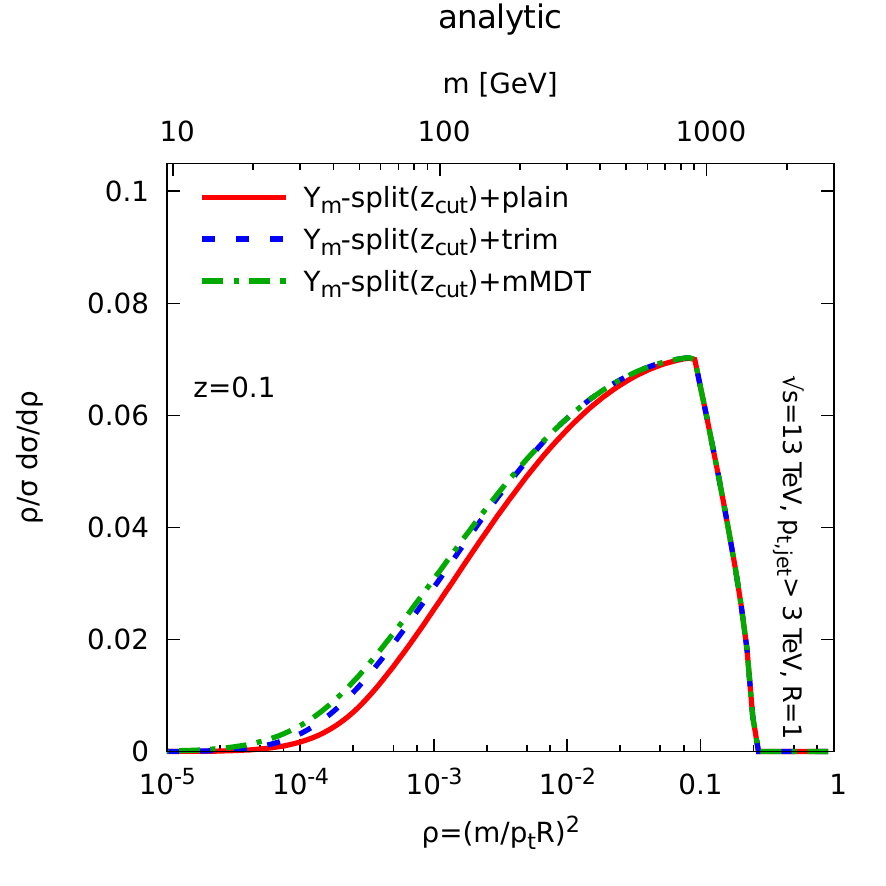}%
  }
  \caption{Mass distributions obtained after imposing a \Ym-splitter
    condition with a $\zcut$ instead of a $\ycut$, followed by an
    optional grooming (mMDT or trimming) step. The plot compares our
    analytic results including multiple-emission corrections (right)
    to Pythia simulations (left).}\label{fig:zcut-mcplot}
\end{figure}

A comparison between (\ref{eq:ysplitter-zcut-with-me}) and
Monte-Carlo simulations is provided in Fig.~\ref{fig:zcut-mcplot}.

Despite the simplicity of the analytic results, and the fact that the
general shape is well reproduced by the analytic results, one should
note that the Monte-Carlo simulations show a slightly larger spread
between the different groomers than what was observed with a $\ycut$
\Ym-splitter condition, indicating a larger impact of subleading terms
for the $\zcut$ condition.
A complete calculation at the single-logarithmic accuracy would
however require the inclusion of several additional effects like
soft-and-large-angle emissions, 2-loop corrections to the running of
the strong coupling and non-global logarithms.

Furthermore, the mass spectrum is slightly higher at small masses with
a $\zcut$ than with a $\ycut$, and we should therefore expect a
slightly better tagging performance for the latter.
This can be seen directly in the Monte-Carlo plots in
Figs.~\ref{fig:mass-ordered-mcplot} and \ref{fig:zcut-mcplot}, and 
ought to be apparent from an analytic calculation including multiple emissions
also for the $\ycut$ case.
Physically, we attribute that to the fact that the \Ym-splitter
condition including multiple emissions is more constraining in the case
of a $\ycut$, Eq.~(\ref{eq:Ym-split-multiple-emissions}), than with a
$\zcut$, Eq.~(\ref{eq:Ym-zcut-multiple-emissions}).

Conversely, as was already observed for a $\zcut$-based compared to
a $\ycut$-based mMDT \cite{Dasgupta:2013ihk}, one should expect a
$\zcut$-based \Ym-splitter to be less sensitive to non-perturbative
effects than a $\ycut$-based \Ym-splitter.
We will confirm this in our study of non-perturbative effects in
section~\ref{sec:np-effects}.

\subsection{Y-splitter with SoftDrop pre-grooming}\label{sec:sd}

\begin{figure}
 \begin{minipage}[c]{0.45\textwidth}
    \includegraphics[width=\textwidth]{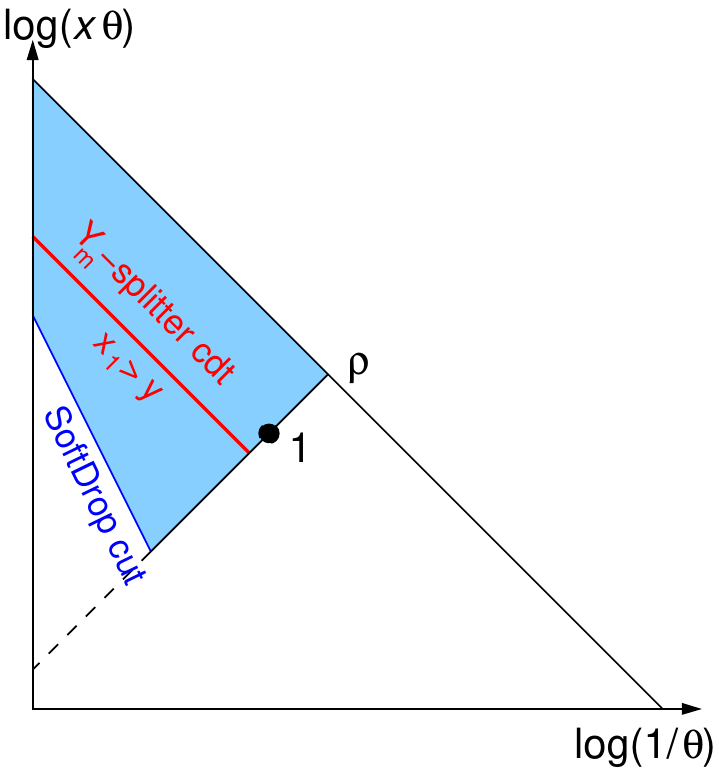}
  \end{minipage}\hfill
  \begin{minipage}[l]{0.4\textwidth}
    \caption{Lund diagram corresponding to Y-splitter applied on a
      pre-groomed jet with SoftDrop. The shadowed area corresponds to
      the region allowed by SoftDrop and entering into the Sudakov
      factor.
      The dashed (red) line corresponds to the \Ym-splitter condition.
    }\label{fig:sd-lund}
  \end{minipage}
\end{figure}

There is one last possible adaptation of the Y-splitter method that we
wish to introduce. 
Our original motivation to combine Y-splitter with grooming was to
reduce the sensitivity of the plain jet mass to non-perturbative
effects, especially important for the consequent loss of signal efficiency.
We have then considered the mMDT  and trimming as
possible ways to solve that issue.
For these situations, we have shown that it was crucial to apply the
Y-splitter condition on the plain jet mass and use grooming to
determine the final jet mass after applying the Y-splitter condition.

There is however an alternative, and in some sense intermediate,
possibility. Instead of using the modified MassDropTagger or trimming
we can groom the jet using SoftDrop~\cite{Larkoski:2014wba}. 
More precisely, one first applies a SoftDrop procedure --- with
parameters $\zetacut<\ycut$ and $\beta$ --- to the jet in order to
reduce the non-perturbative effects and, after this pre-grooming step,
we impose the Y-splitter condition on the pre-groomed jet.

In practice, this would be very similar to the case of the plain jet
mass discussed in section~\ref{sec:pure} except that it would apply
to a SoftDropped jet in which soft and large-angle emissions have been
groomed away.
Focusing on the \Ym-splitter case, \ie using a mass declustering, it
is straightforward to realise that the mass distribution would be
given by
\begin{equation}\label{eq:preSD-resum}
\frac{\rho}{\sigma} \frac{d\sigma}{d\rho}^{\text{LL}} = 
\int_{\frac{y}{1+y}}^{\frac{1}{1+y}}dx_1\,P(x_1)\,\frac{\alpha_s(x_1\rho)}{2\pi}
e^{-R_{\text{SD}}(\rho)} ,
\end{equation}
where the Sudakov exponent, graphically represented in
Fig.~\ref{fig:sd-lund}, now includes the effect of SoftDrop
\begin{equation}
R_{\text{SD}}(\rho) = \int \frac{d\theta^2}{\theta^2}dx\,P(x)\,
  \frac{\alpha_s(x^2 \theta^2)}{2\pi}
  \,\Theta \left(x > \zetacut \theta^\beta \right)
  \,\Theta \left(x \theta^2>\rho \right).
\end{equation}

As for the ``pure'' \Ym-splitter case discussed in
section~\ref{sec:mass-ordering}, this result captures the leading
behaviour, without any additional subleading logarithms of
$\ycut$ to resum.
Furthermore, (\ref{eq:preSD-resum}) is also largely unaffected by
a possible mMDT or trimming one would apply after the \Ym-splitter
condition since the latter guarantees that the emission that dominates
the mass carries a momentum fraction larger than
$\ycut$.\footnote{Differences between groomers would still apply due
  to sub-leading single logarithmic terms coming from multiple-emission
  contributions to the jet mass. Note also that in the case of
  trimming, there would be an interference between the SoftDrop and
  trimming conditions when the latter starts cutting angles smaller
  than $\Rtrim$, which occurs for $\rho=\zetacut \Rtrim^{2+\beta}$.}

\begin{figure}
  \centerline{%
    \includegraphics[width=0.49\textwidth]{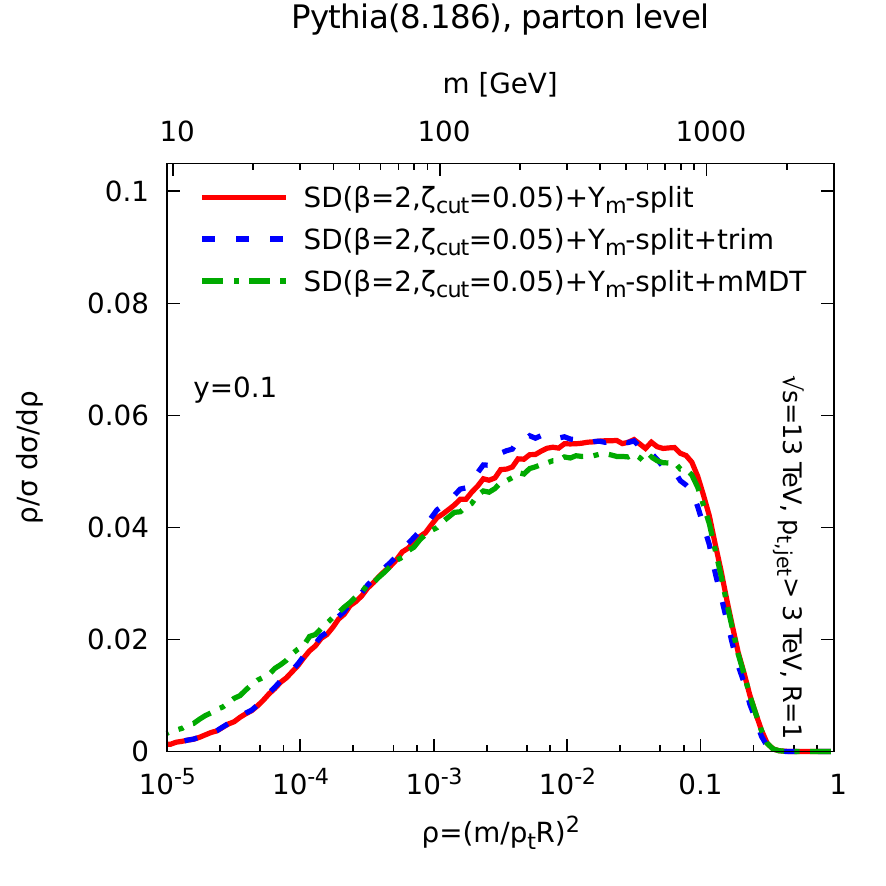}%
    \hfill
    \includegraphics[width=0.49\textwidth]{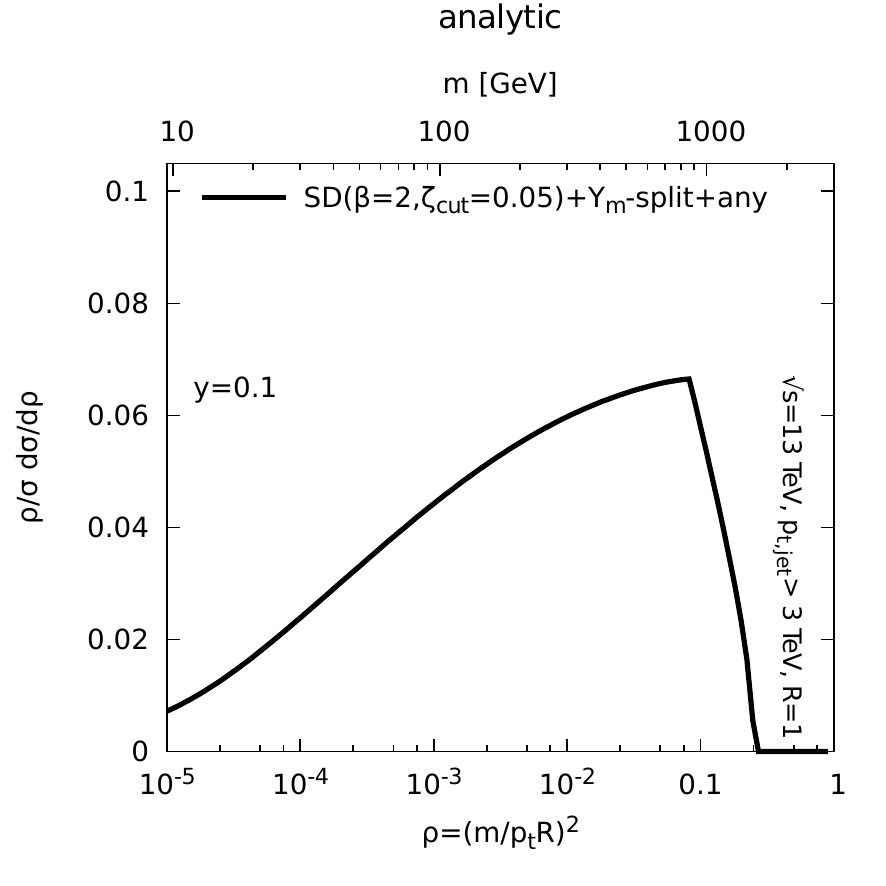}%
  }
  \caption{The solid (red) curve on the left plot shows the mass
    distribution obtained with Pythia8 by first applying a SoftDrop pre-grooming
    with $\zeta_{\text{cut}}=0.05$ and $\beta=2$ and then imposing the
    \Ym-splitter condition $\ycut>y$. For the dashed and dash-dotted
    lines on the left plot we have then applied an extra grooming step
    (trimming and the mMDT, respectively).
    The right plot shows the corresponding LL analytic prediction
    (\ref{eq:preSD-resum}) which is common to all three
    setups. }\label{fig:sd-mcplot}
\end{figure}

Compared to the pure Y-splitter case,
Eq.~(\ref{eq:pure-resum-leading}), we should expect the pre-groomed
result (\ref{eq:preSD-resum}) to show a worse performance. This is due
to the fact that SoftDrop grooms away a region of the phase-space that
would otherwise be constrained in the ungroomed case, resulting into a
smaller Sudakov suppression for the SoftDrop+Y-splitter case compared
to the pure Y-splitter case.
Conversely, the region which is groomed away is also the region which
is expected to be the most affected by non-perturbative effects, the
Underlying Event in particular. We should therefore expect the
pre-groomed Y-splitter to be more robust against non-perturbative
effects. This will be made explicit in the next section.

Note also that, although we have advocated so far that it is important
to apply the groomer after the Y-splitter condition, here we apply the
grooming procedure first.
This makes sense since we here apply a much gentle grooming procedure
--- SoftDrop with positive $\beta$ --- and, as a consequence, we still
benefit from a large Sudakov suppression.

Finally, we have compared our analytic result~(\ref{eq:preSD-resum})
with Pythia8 Monte-Carlo simulations in Fig.~\ref{fig:sd-mcplot} and
we see once again that it does capture the overall behaviour. We also
notice in the Monte-Carlo simulations that once the pre-grooming step
has been applied, the effect of an extra grooming (mMDT or trimming)
has almost no effect.

\section{Non-perturbative effects}\label{sec:np-effects}

Our discussion has so far focused on pure perturbative effects.
It is nevertheless also important to assess the size of
non-perturbative effects, which we would like to be as small as
possible, for better theoretical control.

To estimate non-perturbative effects, we have used Pythia8 with tune
4C~\cite{Corke:2010yf} to simulate $W$ jets (our signal, obtained from
$WW$ events) and quark jets (our background, obtained from $qq\to qq$
Born-level events). For each event, we select the (plain) jets passing
a given $p_t$ cut that we shall vary between 250~GeV and 3~TeV and
then apply one of the tagging procedures used in this paper to obtain
a mass distribution for the signal and background jets. For
Y-splitter, we have used a $\ycut$ (or $\zcut$) of 0.1, adapting the
mMDT and trimming energy cut accordingly.
Finally, in order to obtain the signal and background efficiencies we
have kept jets which, after the whole procedure, have a mass between
60 and 100~GeV.
All efficiencies presented in this section are normalised to the total
inclusive jet cross-section to obtain ($W$ or quark) jets above the
given $p_t$ cut.

\begin{figure}
  \centerline{%
    \includegraphics[width=0.49\textwidth]{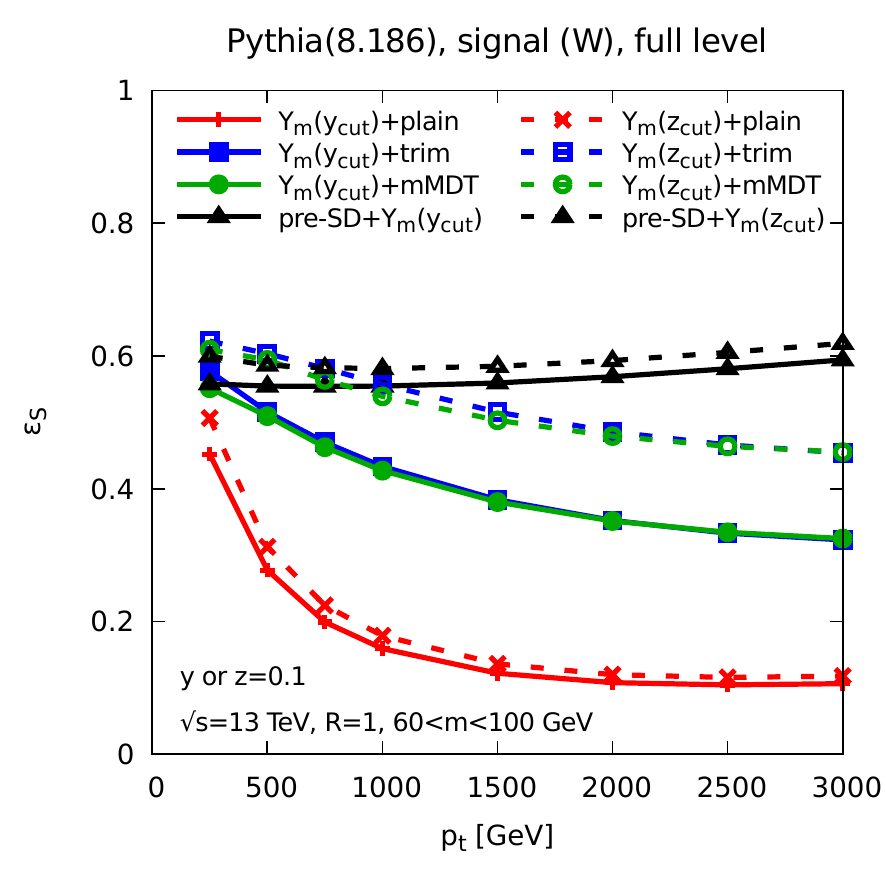}%
    \hfill
    \includegraphics[width=0.49\textwidth]{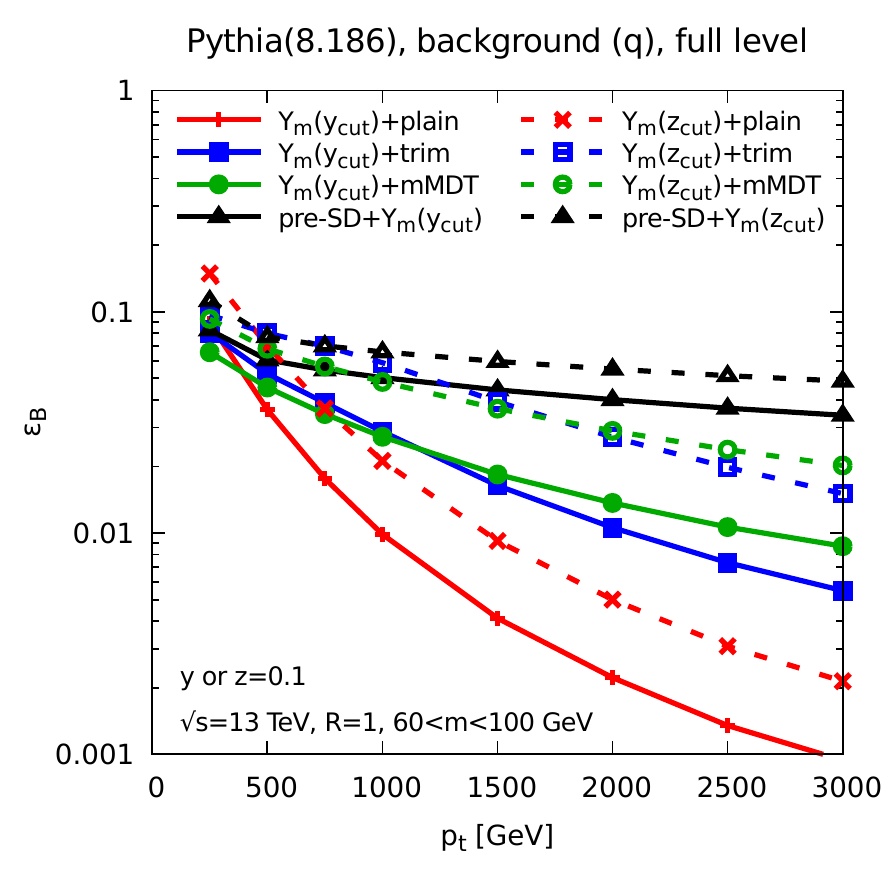}%
  }
  \caption{Signal and background efficiencies for a few selected
    tagging methods. the left-hand plot corresponds to signal ($W$
    jets) and the right-hand plot to background (quark) jets. For both
    plots, full events, including hadronisation and the Underlying
    Event, have been used. Different point types
    (and colours) correspond to different grooming (or pre-grooming)
    methods; solid (resp. dashed) lines are obtained applying a
    \Ym-splitter $\ycut$ (resp. $\zcut$) condition. Details are given
    in the main text.}\label{fig:eff-v-pt-SandB}
\end{figure}

\begin{figure}
  \centerline{%
    \includegraphics[width=0.49\textwidth]{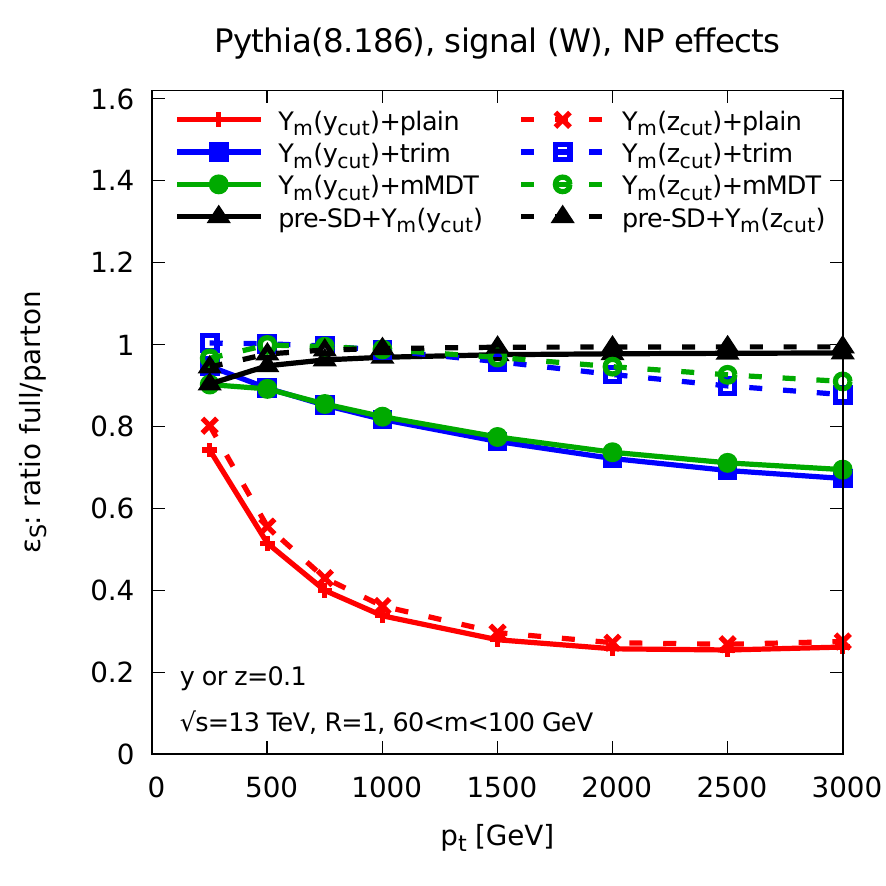}%
    \hfill
    \includegraphics[width=0.49\textwidth]{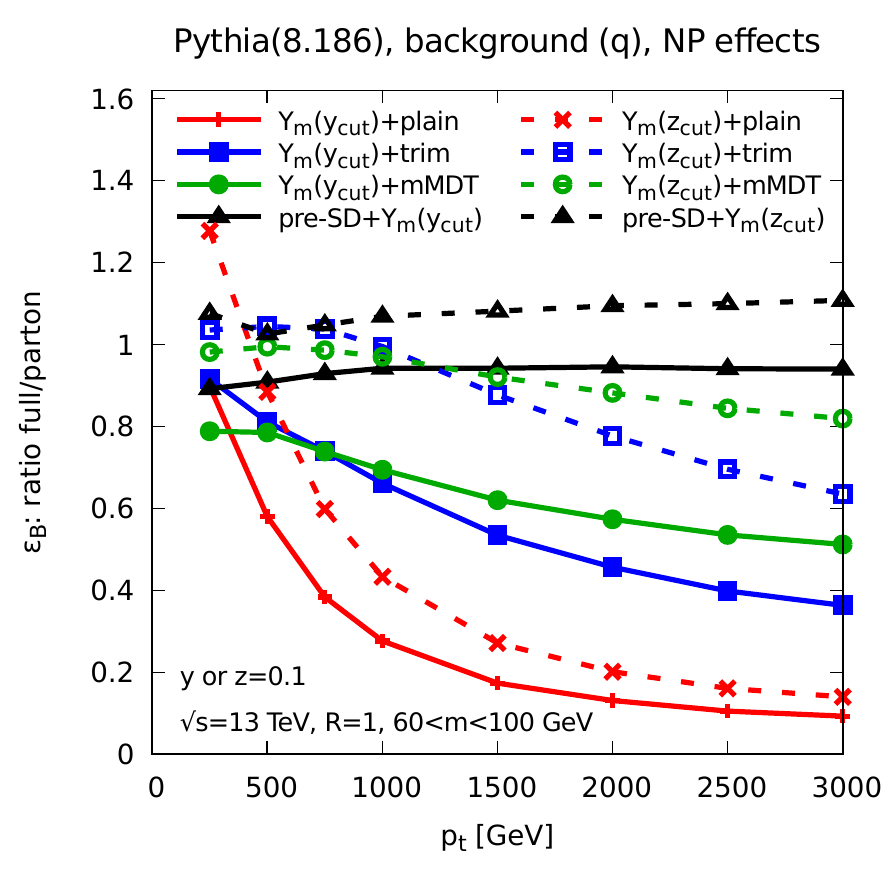}%
  }
  \caption{Non-perturbative corrections for signal (left) and
    background (right) efficiencies due to hadronisation and the
    Underlying Event, computed as a ratio of efficiencies obtained
    with and without non-perturbative effects. Different point types
    (and colours) correspond to different grooming (or pre-grooming)
    methods; solid (resp. dashed) lines are obtained applying a
    \Ym-splitter $\ycut$ (resp. $\zcut$) condition. Details are given
    in the main text.}\label{fig:eff-v-pt-SandB-np}
\end{figure}

Throughout this paper, we have considered a large range of Y-splitter
conditions ($k_t$ or mass declustering, $\ycut$ or $\zcut$) and
grooming options (ungroomed jets, mMDT, trimming or pre-grooming).
It is hopeless to compare all possible combinations in a
human-readable plot.
We have therefore selected a few representative cases to illustrate
both signal-v-background performance and sensitivity to
non-perturbative effects.
Between Y-splitter and \Ym-splitter conditions, we have limited
ourselves to the latter, since it has a slightly better performance
than the former.\footnote{The better performance is expected from our
  analytic calculations and also confirmed directly in
  Monte-Carlo-based studies.}
We have considered both a $\ycut$ and a $\zcut$ type of condition,
using in practice $\ycut=\zcut=0.1$.
We have then studied 4 grooming options: the ungroomed (or pure)
case which acts as a baseline, mMDT and trimming both applied after
the \Ym-splitter condition, and SoftDrop pre-grooming for which the
\Ym-splitter condition is applied after the pre-grooming.
With a $\ycut$-based \Ym -splitter condition, the momentum fraction
used in the mMDT and trimming is set to $\ycut/(1+\ycut)$, while for a
$\zcut$-based \Ym-Splitter condition it is simply set to $\zcut$.
For the SoftDrop pre-grooming, we have set $\beta=2$ and
$\zetacut=0.05$.

The signal and background efficiencies obtained from our simulations
when varying the boosted jet $p_t$ are presented in
Fig.~\ref{fig:eff-v-pt-SandB} for simulations including hadronisation
and the Underlying Event. This should be considered together with
Fig.~\ref{fig:eff-v-pt-SandB-np} where we have plotted the ratio of
the efficiencies obtained with hadronisation and the Underlying Event
to those obtained without, as a measure of non-perturbative effects.

\begin{figure}
  \begin{minipage}[c]{0.49\textwidth}
    \includegraphics[width=\textwidth]{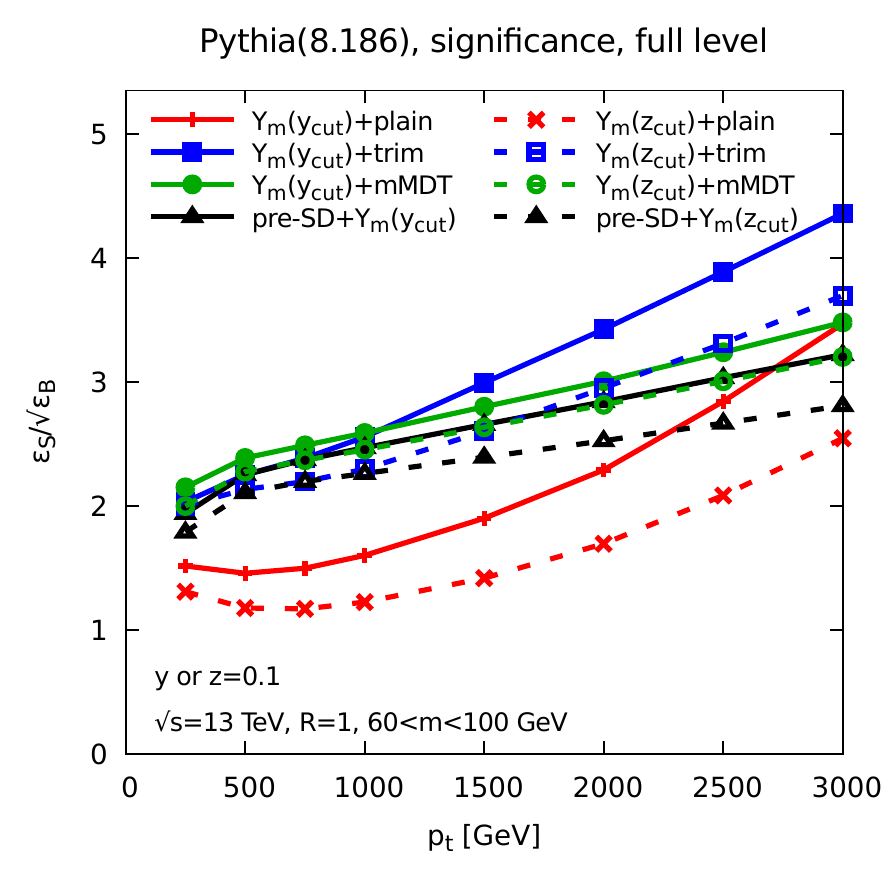}%
  \end{minipage}\hfill
  \begin{minipage}[l]{0.4\textwidth}
    \caption{Signal significance obtained from the efficiencies in
      Fig.~\ref{fig:eff-v-pt-SandB}. Again, both hadronisation and the
      Underlying Event are included. Different point types (and
      colours) correspond to different grooming (or pre-grooming)
      methods; solid (resp. dashed) lines are obtained applying a
      \Ym-splitter $\ycut$ (resp. $\zcut$) condition. Details are
      given in the main text.}\label{fig:eff-v-pt-SsqrtB}
  \end{minipage}
\end{figure}

For a more direct comparison of the performance of the variants of
Y-splitter we have considered here, we have shown the resulting signal
significance, computed as $\varepsilon_S/\sqrt{\varepsilon_B}$ in
Fig.~\ref{fig:eff-v-pt-SsqrtB} which again, has to be considered
together with the size of non-perturbative effects shown in
Fig.~\ref{fig:eff-v-pt-SandB-np}.

Based on this series of plots, we can make several observations. 
First, for the plain jet mass case with either Y-splitter option, we
see that both the signal and background efficiencies are lower than
for the groomed cases. Such a large difference is in part due to the
much larger sensitivity to the non-perturbative effects, our initial
motivation to investigate the combination of Y-splitter with grooming
techniques.

Next, we had noticed in sections \ref{sec:zcut} and \ref{sec:sd},
based on our analytic calculations, that if instead of imposing a
\Ym-splitter condition computed on the plain jet with a $\ycut$, we
were either imposing a $\zcut$ condition or pre-grooming the jet with
SoftDrop, it would translate to a larger $\epsilon_B$.
This is indeed confirmed by these Monte-Carlo simulations.

Furthermore, we also observe large differences in terms of the various
sensitivities to non-perturbative effects.
Compared to the pure Y-splitter case, applying grooming (either
trimming or mMDT) reduces the sensitivity to non-perturbative effects,
with the mMDT being slightly less sensitive than trimming (albeit also
with a slightly smaller discriminative power as indicated by the
signal significance). 

The same observation can be made about the use of a pre-grooming
procedure before computing \Ym-splitter: the background suppression is
clearly less pronounced than for all the other cases considered here,
but it only leads to ~10\% non-perturbative corrections whereas in the
case of \Ym-splitter+trimming, which gives the best performance,
non-perturbative effects reach ~60\%.

We should stress that when a given method suppresses the background
more than another, it also tends to reduce the signal more. It is
therefore far from obvious that a larger background suppression would
ultimately lead to a larger significance,
$\varepsilon_S/\sqrt{\varepsilon_B}$.
However, differences observed in background efficiencies are usually
exponential --- notice the logarithmic scale on the right-hand plot of
Fig.~\ref{fig:eff-v-pt-SandB} --- and are therefore expected to have
more impact than smaller variations in signal efficiencies.
The ordering is therefore usually respected when we look at the signal
significance, Fig.~\ref{fig:eff-v-pt-SsqrtB}.

\section{Discussion and Conclusions}

In this paper, we have studied analytically the effect of imposing a
Y-splitter condition on boosted jets.
Based on previous work~\cite{Dasgupta:2015yua} which had shown good
performance in Monte-Carlo simulations, we have considered the
combination of a Y-splitter cut together with a grooming procedure.
Specifically we have studied the impact of trimming and the modified
MassDropTagger which act here as groomers \ie serve to
limit the impact of non-perturbative effects on the jet. It is the
Y-splitter condition which plays the role of the tagger, and hence reduces the
QCD background.

We have also considered variants of the Y-splitter condition:
first the standard one defined in terms of a cut  on $k_t^2/m^2$
(known also as a $\ycut$ condition), secondly a
variant called \Ym-splitter where the $k_t$ scale is computed using a
``mass declustering'', \ie by undoing the last step of a
generalised-$k_t$ clustering with $p=1/2$, and finally replacing the
standard $\ycut$ condition by a $\zcut$ condition,
Eq.~(\ref{eq:ysplitter-cdt-zcut}), where we cut directly on the subjet
momentum fractions instead of $k_t^2/m^2$.
For each variant, we then study different combinations with and
without grooming. Specifically, imposing the Y-splitter condition on
the plain jet we examine the jet mass without any grooming
(``Y+plain'') or perform subsequent grooming and study either the
trimmed jet mass (``Y+trim'') or the mMDT jet mass (``Y+mMDT'').
Alternatively, we can apply a more gentle SoftDrop grooming to the
jet and then impose the Y-splitter condition and compute the jet mass
on that pre-groomed jet (``SD+Y'').

The main result of the paper is that, keeping only the dominant terms
enhanced by logarithms of the jet mass at all orders (LL), the same
behaviour is recovered for all these variants when applied to QCD
background jets. It is given by
Eq.~(\ref{eq:pure-resum-leading}) or Eq.~(\ref{eq:preSD-resum}) when
the Y-splitter condition is computed on the plain jet or the SD jet,
respectively.
Furthermore, for QCD jets applying a grooming procedure to compute
the jet mass after imposing the Y-splitter condition only brings
subleading corrections, and thus its main role is to ensure a decent
resolution when measuring the jet mass by reducing the non-perturbative
and pileup effects.

Technically, the good performance of the Y-splitter+grooming boosted
object tagger comes from the combination of two effects. Firstly  for
the pure Y-splitter case (i.e. without grooming) the QCD background is
suppressed relative to the case of the plain jet mass. One obtains an
exponential Sudakov factor, double-logarithmic in the jet mass, which is then multiplied
by a prefactor containing a modest logarithm  in $\ycut$, \ie smaller than for
the plain jet mass where the prefactor has instead a logarithm
involving $m/p_t$. Secondly the use of grooming does not significantly affect
this background suppression due to the fact that it induces only
subleading corrections to the pure Y-splitter case. On the other hand
the use of grooming considerably improves the signal efficiency
relative to the pure Y-splitter case.

Further, if one considers in more detail the role of subleading corrections
induced by grooming we have seen that they only introduce numerically
modest differences between the various methods we have
considered. While these differences are clearly visible in both analytical and
Monte Carlo studies, their size is insufficient to radically alter the
performance of the the tagger. In some cases we have shown
that including a  resummation of all the double-logarithmic terms
(LL+LL$_y$), either in the jet mass or in $\ycut$, captures the main
characteristics of these differences.
Monte-Carlo simulations also confirm that all the Y-splitter variants we
have considered are to a large extent compatible with
Eq.~(\ref{eq:pure-resum-leading}).

\begin{figure}
  \centering
  \includegraphics[width=0.8\textwidth]{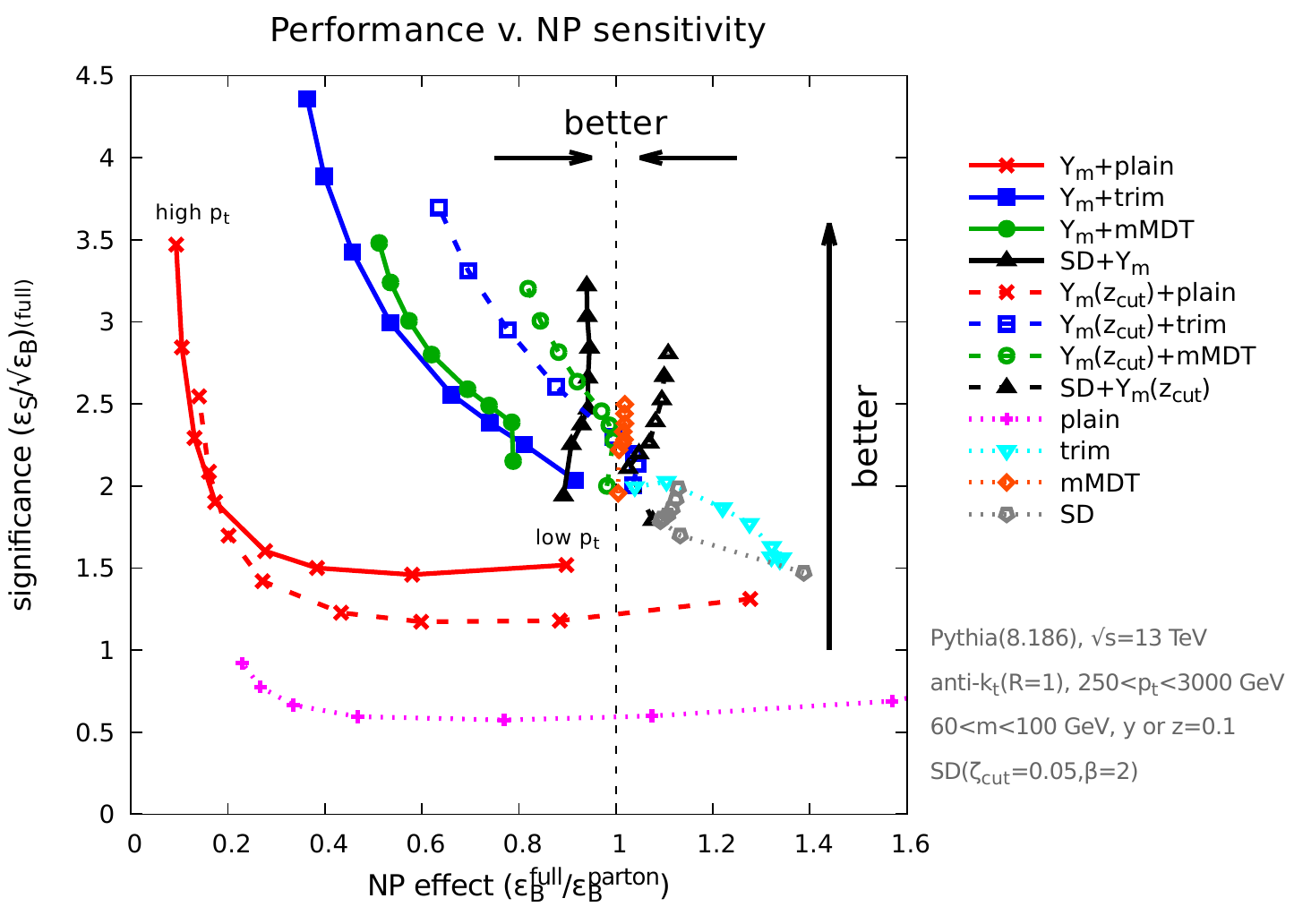}
  \caption{Summary plot showing the signal efficiency, computed as
    $\epsilon_S/\sqrt{\epsilon_B}$ for events at particle (full)
    level, versus the corresponding size of non-perturbative
    effects, estimated by the ratio of the background efficiency
    calculated, for a quark-jet sample, at particle (full) level and
    at parton level.
    The different points on each curve correspond to different values
    of the jet $p_t$, spanning from 250~GeV to 3~TeV.
    Each curve represents a specific method. We show the two
    variants of \Ym-splitter, either with a standard $\ycut$ condition
    (solid lines) or with a $\zcut$ condition (dashed lines, see
    Eq.~(\ref{eq:ysplitter-cdt-zcut})), with $\ycut=\zcut=0.1$.
    Results are presented for a \Ym-splitter condition computed on the
    plain jet followed by a computation of either the plain jet mass
    (red), the trimmed jet mass (blue) or the mMDT jet mass
    (green). For the black curve, we have computed both the
    \Ym-splitter condition and the mass on a SoftDropped jet with
    $\beta=2$ and $\zeta_{\text{cut}}=0.05$.
    Finally, we also added for comparison the results obtained
    without the Y-splitter condition for either the plain jet mass or
    the groomed jet mass.
    In all cases, we have required that the mass is between 60 and
    100~GeV, and signal and background efficiencies are computed wrt
    the inclusive jet rate for each $p_t$ cut.  }\label{fig:summary}
\end{figure}

In order to discuss in detail the physical properties of all these variants and
compare them, several criteria have to be considered.
To facilitate the discussion, we have considered the Monte-Carlo setup
described in section~\ref{sec:np-effects} and have plotted in
Fig.~\ref{fig:summary} two important quantities when considering the
performance of a boosted-object tagging method: on the vertical axis
we show the raw performance of the method, measured as usual by the signal
significance. On the horizontal axis we have a measure of
the method's robustness defined in terms of insensitivity to
non-perturbative contributions. Here we have used a
non-perturbative correction factor defined as the
ratio of the efficiencies at particle (full) and parton levels and
have explicitly considered the case of quark jets, with similar trends
expected for gluon jets.
Ideally, we want a method with high performance and robustness, \ie
with a large signal significance and a non-perturbative correction
factor close to 1.
We can then make the following generic observations:
\begin{itemize}
\item {\it Effect of grooming.}
  It is obvious from Fig.~\ref{fig:summary} that adding grooming
  improves considerably both the performance and the robustness. Based
  on what we have discussed before, the improvement in performance
  comes mainly from the impact on signal efficiency.
  However it is crucial to impose the Y-splitter constraint on the
  plain jet instead of the groomed jet, otherwise one only gets a much
  smaller Sudakov suppression of the QCD background.\footnote{In that
    case, one recovers a Sudakov similar to that of the groomer, which
    is much smaller than the plain jet mass Sudakov, see
    Appendix~\ref{app:grooming+Ysplitter} for an explicit example.}
  We should however stress that subleading corrections sometimes come
  with several transition points in the mass distribution, which can
  be an issue for practical applications in an experimental context.
\item {\it $k_t$ or mass declustering?} 
  As we have seen in our calculations, even though they lead to the
  same LL result, the overall analytic structure is found to be much
  simpler for the case of mass declustering. In particular, the
  groomed (trimmed or mMDT) and plain jet results are given by the LL
  result with no additional double-logarithmic contributions in the
  LL+LL$_y$ approximation. Corrections to that result would be
  purely single-logarithmic in the jet mass, \eg coming from multiple
  emissions.
  Then, although it is not explicitly shown in the figure, using mass
  declustering comes with a small gain in performance. We traced it
  back to the absence of the extra terms between the LL and LL+LL$_y$
  results.
\item {\it Trimming or mMDT?}
  At LL accuracy, both give the same perturbative performance.
  In practice,  at large $p_t$ we see that trimming tends
  to give a slightly better performance and is slightly less
  robust.
  It remains to be investigated whether this is generally true or a
  consequence of our specific choice of parameters (see ``A word of
  caution'' below).
  Even if it was a general  observation, it is not obvious that one
  should prefer trimming over the mMDT. Indeed, we have seen that
  trimming introduces more transition points (and therefore kinks)
  in the mass distribution than the mMDT, although they are reduced by
  the use of \Ym-splitter). These can have undesirable effects in
  experimental analyses, \eg for side-band estimates of the
  backgrounds or if the signal lies on top of a transition
  point. 
\item {\it $\ycut$ or $\zcut$?}
  Contrary to the case of $k_t$ v. mass declustering, the situation is
  less obvious here: the $\ycut$ variant shows a better performance,
  in part traced back to single-logarithmic effects like multiple
  emissions, but at the same time the $\zcut$ variant appears less
  sensitive to non-perturbative effects.
  The choice between the two is therefore again a trade-off between
  performance and robustness.
  In terms of the analytic structure of the results, we should point
  out that the $\zcut$ variant is likely more amenable to a higher
  logarithmic accuracy resummation more than the $\ycut$ version. In
  particular it gives a simple expression for the resummation of
  multiple emission effects.
\item {\it Pre-grooming.}
  We see yet again the same trade-off between performance which is globally in
  favour of \Ym-splitter+grooming, and robustness which is globally in favour
  of pre-grooming. 
  The differences in performance are explicitly predicted by our
  analytic results, already at LL accuracy.
  The differences in robustness are also expected from the fact that
  Soft-Drop cuts out soft-and-large-angle radiation.
  It is however interesting to notice that compared to the results
  obtained for mMDT, trimming and SoftDrop alone, the addition of the
  \Ym-splitter condition still results in a sizeable performance gain.
\item {\it A word of caution.}
  We should point out that Fig.~\ref{fig:summary} was obtained for one
  specific choice of the free parameters like the jet radius, $\ycut$,
  $\zcut$ or mass-window parameters. 
  In practice, we do not expect to see substantial differences if we
  were to adopt a different setup, especially for the main features
  which are backed  up by analytic calculations.
  However, some of the differences observed in Fig.~\ref{fig:summary}
  go beyond our analytic accuracy and can depend on our choice of
  parameters. This concerns, in particular, the subleading differences
  observed between trimming and the mMDT, or details about the precise
  size of non-perturbative effects.
\end{itemize}

In summary we advocate the use of Y-splitter with grooming as a
superior boosted object tagger for hadronic two-body decays, as was
first noted in Ref.~\cite{Dasgupta:2015yua}. While this initial
observation was based on Monte Carlo studies alone, in the present
paper we have put it on much firmer ground via adding an analytical
first principles (i.e. model independent) understanding of the results
for QCD background jets. We have also investigated several variants
both by using different grooming methods as well as by modifying the
standard Y-splitter algorithm in various ways. Eventually the results
for different variants indicate that there is a trade-off between
performance and robustness.
Such a trade-off was also observed in the case of jet
shapes~\cite{Dasgupta:2015lxh} where the addition of grooming also
resulted in smaller sensitivity to non-perturbative effects at the
expense of discriminating power.
In terms of sheer performance as reflected by the signal significance,
the \Ym-splitter+trimming or \Ym-splitter+mMDT combinations with a
standard $\ycut$ should be preferred.
If instead we want maximum robustness, \eg to reduce uncertainties,
\Ym-splitter+mMDT with a $\zcut$ condition or SoftDrop pre-grooming
(with either a $\ycut$ or a $\zcut$ condition) appear at the same time
both efficient and robust.
Indeed, these variants still outperform the standard methods such as
pure mMDT, pure trimming or pure SoftDrop at high $p_t$ as is evident
from Fig.~\ref{fig:summary} .

For the combinations which show a small sensitivity to
non-perturbative effects, it would be interesting to push the
analytic calculations beyond the precision targeted in this paper.

Also, it remains to optimise the parameters of the tagger in order to
maximise the performance which we leave to forthcoming work.

Lastly, it remains to be determined as to whether declustering using
the generalised-$k_t$ algorithm with $p=1/2$ yields the best
performance. In that respect it would be interesting to study smaller
values of $p$.\footnote{See \eg Appendix~C of
  Ref.~\cite{Dasgupta:2015lxh}.}

\section*{Acknowledgements}
We are all grateful to Gavin Salam and Andrzej Siodmok for helpful
discussions.
MD and GS both thank each other's institutions for hospitality as this
work was being carried out.
MD's work is supported in part by the Lancaster-Manchester-Sheffield
Consortium for Fundamental Physics under STFC grant ST/L000520/1.
GS's work is supported in part by the French Agence Nationale de la
Recherche, under grant ANR-15-CE31-0016 and by the ERC advanced grant
Higgs@LHC.

\appendix

\section{Radiators and friends}\label{app:radiators}

In this appendix, we give explicit expressions for the various
radiators that appeared throughout this paper.

The running coupling scale runs according to
\begin{equation}
\alpha_s(k_t^2) = \frac{\alpha_s}{1-2\alpha_s\beta_0\ln(p_tR/k_t)},
\end{equation}
where $\alpha_s$ is taken at the scale $p_tR$ and
$\beta_0=(11C_A-4n_fT_R)/(12\pi)$.
To avoid hitting the Landau pole, the coupling is frozen at
$k_t=\mu_{\text{fr}}$.

We consider a jet of a given flavour with colour factor $C_R$ ($C_F$
for quark jets and $C_A$ for gluon jets) and hard-splitting constant
$B_i$ with
\begin{equation}
B_q = -\frac{3}{4}\qquad\text{and}\qquad B_g=-\frac{11C_A-4n_fT_R}{12C_A}.
\end{equation}

For convenience, it is helpful to define
\begin{align}
L_\rho & = \ln(1/\rho), & L_y   & = \ln(1/y),\\
L_r&=\ln(1/\Rtrim^2),   &  L_c&=\ln(1/\zcut^2),\\
L_{\text{fr}} & = \ln(1/\tilde\mu_{\text{fr}}), & L_{k_t} & = \ln(1/k_t),
\end{align}
with $\tilde\mu_{\text{fr}}=\mu_{\text{fr}}/(p_tR)$. For any $x$ in
one of the above logarithms, we also introduce the short-hand
notation,
\begin{equation}
\lambda_x = 2\alpha_s\beta_0L_x,
\end{equation}
and use $W(x)=x\,\ln x$.

All the radiators in this paper can be easily expressed in terms of
a single generic construct. Let us consider two $k_t$ scales $k_{t0}$
and $k_{t1}<k_{t0}$, and a parameter $\alpha\ge 0$. We then define
$k_{t2}=(k_{t0}k_{t1}^{1+\alpha})^{1/(2+\alpha)}$, 
$L_i=\ln(1/k_{ti})$ and $\lambda_i=2\alpha_s\beta_0L_i$. The basic
quality of interest can be written as 
\begin{align}
T_{\alpha}(L_0, L_1)
 & \mathrel{\makebox[\widthof{a+=x}]{$=$}}
  \int\frac{d\theta^2}{\theta^2}\,\frac{dx}{x}\,\frac{\alpha_s(x^2\theta^2)}{2\pi}\,\Theta(x\theta^2>k_{t1})\,\Theta(x<k_{t0}\theta^\alpha)
  \\
 & \mathrel{\makebox[\widthof{a+=x}]{$\overset{L_1<L_{\text{fr}}}{=}$}}
   \frac{C_R}{2\pi\alpha_s\beta_0^2}\bigg[
     \frac{W(1-\lambda_0)}{1+\alpha}
    -\frac{2+\alpha}{1+\alpha}W(1-\lambda_2)
    +W(1-\lambda_1)\bigg]\nonumber\\
 & \mathrel{\makebox[\widthof{a+=x}]{$\underset{L_2<L_{\text{fr}}}{\overset{L_1>L_{\text{fr}}}{=}}$}}
   \frac{C_R}{2\pi\alpha_s\beta_0^2}\bigg[
     \frac{W(1-\lambda_0)}{1+\alpha}
    -\frac{2+\alpha}{1+\alpha}W(1-\lambda_2)
    +(1-\lambda_1)\ln(1-\lambda_{\text{fr}})+\lambda_{\text{fr}}-\lambda_1\bigg]\nonumber\\
 &\mathrel{\makebox[\widthof{a+=x}]{}}
  +\frac{\alpha_s(\tilde\mu_{\text{fr}}^2)C_R}{\pi}(L_1-L_{\text{fr}})^2 \nonumber\\
 & \mathrel{\makebox[\widthof{a+=x}]{$\underset{L_0<L_{\text{fr}}}{\overset{L_2>L_{\text{fr}}}{=}}$}}
   \frac{C_R}{2\pi\alpha_s\beta_0^2}\bigg[
     \frac{1-\lambda_0}{1+\alpha}\ln\Big(\frac{1-\lambda_0}{1-\lambda_{\text{fr}}}\Big)
    +\frac{\lambda_0-\lambda_{\text{fr}}}{1+\alpha}\bigg]\nonumber\\
 &\mathrel{\makebox[\widthof{a+=x}]{}}
  +\frac{\alpha_s(\tilde\mu_{\text{fr}}^2)C_R}{\pi}\bigg[(L_1-L_2)^2
       + \frac{L_2-L_{\text{fr}}}{1+\alpha}(L_2+L_{\text{fr}}-2L_0)\bigg] \nonumber\\
 & \mathrel{\makebox[\widthof{a+=x}]{$\overset{L_0>L_{\text{fr}}}{=}$}}
   \frac{\alpha_s(\tilde\mu_{\text{fr}}^2)C_R}{\pi}\frac{1}{2+\alpha}(L_1-L_0)^2.\nonumber
\end{align}
Note that we tacitly assume that $T_\alpha(L_0,L_1)=0$ if $L_0>L_1$.

With this at hand, we can express all the radiators in this paper in a
fairly concise form.
The first radiator we need corresponds to the plain jet mass

\begin{equation}
R_{\text{plain}}(\rho) 
  = \int\frac{d\theta^2}{\theta^2}\,dx\,P_i(x)\,\frac{\alpha_s(x^2\theta^2)}{2\pi}\,\Theta(x\theta^2>\rho)
 = T_0(-B_i,L_\rho).
\end{equation}
Note that compared to standard expressions in the literature, we have
included the contribution from hard collinear splittings, the
``$B_i$'' term, as a (constant) correction to the (logarithm)
arguments in $T_\alpha$. This is equivalent up to subleading terms
proportional to $B_i^2$.
The main advantage of writing $R_{\text{plain}}$ under the above is that both
$R$ and its derivative vanish when $\ln(1/\rho)=-B_i$, providing a
natural endpoint for our distributions. 
Another way of viewing this result is to realise that one can obtain
the contribution from the hard collinear splittings by putting an
upper bound on the $x$ integrations at $x=\exp(B_i)<1$.

Next, we need to specify $R_{k_t}(k_t,\rho)$ appearing e.g. in
(\ref{eq:pure-resum-all-logs}). One easily finds
\begin{equation}
R_{k_t}(k_t,\rho) =
  \int\frac{d\theta^2}{\theta^2}\,dx\,P_i(x)\,\frac{\alpha_s(x^2\theta^2)}{2\pi}\,\Theta(x\theta^2<\rho)\,\Theta(x^2\theta^2>k_t^2)
 = 2T_0\Big(\frac{L_\rho-B_i}{2},L_{k_t}\Big).
\end{equation}

For situations where we use a SoftDrop pre-grooming, we also need to
specify the SoftDrop radiator. Which is readily available from
\cite{Larkoski:2014wba}\footnote{Up to the reabsorption of the $B$
  terms inside the logarithm mentioned above.} 
\begin{equation}
R_{\text{SD}}(\rho) 
 = \int\frac{d\theta^2}{\theta^2}\,dx\,P_i(x)\,
   \frac{\alpha_s(x^2\theta^2)}{2\pi}\,
   \Theta(x\theta^2>\rho)\,\Theta(x>\zcut\theta^\beta)
 = T_0(-B_i,L_\rho)-T_\beta(L_c,L_\rho).
\end{equation}

\section{Why not use the groomed mass in the Y-splitter condition?}\label{app:grooming+Ysplitter}

We have argued in section~\ref{sec:grooming} that we should first
impose the Y-splitter condition on the plain jet and, if the condition
is satisfied, measure the groomed jet mass. 
The motivation to use the groomed jet mass instead of the plain jet
mass is that it significantly reduces the non-perturbative effects,
especially on signal jets, as shown in \cite{Dasgupta:2015yua}.

Given that observation, one might be tempted to also use the groomed
jet mass in the definition of the Y-splitter condition. 
We show in this appendix that this does not lead to an efficient
tagger.

For simplicity, let us use the modified MassDropTagger (trimming
would yield similar results, albeit a bit more complex and involving
additional transition points) and assume that emission 1 dominates the
groomed mass.
We still have two ways to proceed: we can either decluster the groomed
jet or the plain jet to get the $k_t$ scale entering the \Ym-splitter
condition.
The situation where we use the groomed jet is almost trivial: the
declustering will either select emission 1 or an emission, say 2, at
smaller mass and larger $k_t$. In both cases, the resulting Y-splitter
condition is trivially satisfied, since, \eg in the second case,
$k_{t2}^2>k_{t1}^2=x_1\rho>y\rho$. 
Hence, neither the grooming procedure nor the Y-splitter condition
place any constraint on radiation at larger mass in the groomed-away
region, meaning that we would get
\begin{equation}\label{eq:mMDT-then-YS}
\rho \frac{d\sigma}{d\rho} = 
\int_{\frac{y}{1+y}}^{\frac{1}{1+y}}dx_1\,P(x_1)\,\frac{\alpha_s(x_1\rho)}{2\pi} e^{-R_{\text{mMDT}}(\rho)}.
\end{equation}
This has to be compared to Eq.~(\ref{eq:pure-resum-leading}) for the
situation(s), considered in the main text, where we use the plain jet
mass in the \Ym-splitter condition. The result in
(\ref{eq:mMDT-then-YS}) is significantly less efficient since it comes
with a much weaker Sudakov suppression.

Let us assume instead that we decluster the plain jet in order to
define the Y-splitter $k_t$ scale. In the groomed-away region,
emission with $k_t$ smaller than $k_{t1}$ will be
unconstrained. Emission with $k_t$ larger than $k_{t1}$ will also be
allowed since the resulting Y-splitter condition $k_{t2}^2>\rho y$ is
always met due to $k_{t2}^2>k_{t1}^2>\rho y$. We would therefore again
recover (\ref{eq:mMDT-then-YS}).

Finally, let us briefly discuss the case of \Ym-splitter, with mass
declustering applied to the plain jet. This is slightly different
because now there could be an emission, say emission 2, in the
groomed-away region, with a mass larger than $\rho$ and a $k_t$
smaller than $k_{t1}$. In that case the \Ym-splitter condition would
impose $k_{t2}^2>\rho y$, yielding an additional suppression compared
to (\ref{eq:mMDT-then-YS})
\begin{equation}\label{eq:mMDT-then-YmS}
\rho \frac{d\sigma}{d\rho} = 
\int_{\frac{y}{1+y}}^{\frac{1}{1+y}}dx_1\,P(x_1)\,\frac{\alpha_s(x_1\rho)}{2\pi}
   e^{-R_{\text{mMDT}}(\rho)-R_{\text{out,low}-k_t}(\rho)},
\end{equation}
with
\begin{equation}
R_{\text{out,low}-k_t}(\rho) = \int \frac{d\theta^2}{\theta^2}dx\,P(x)\,
  \frac{\alpha_s(x^2 \theta^2)}{2\pi}
  \Theta \left(x \theta^2>\rho \right)\,\Theta(x^2\theta^2<\rho y).
\end{equation}
This is better than (\ref{eq:mMDT-then-YS}) but still remains
less efficient than (\ref{eq:pure-resum-leading}) by double
logarithms of $\rho$.

In the end, it is not our recommendation to use the groomed jet mass
in the Y- or \Ym-splitter condition.

\section{Resummation of the $\ln y$-enhanced terms for Y-splitter
  with the modified MassDrop mass}\label{sec:mMDTYSresum}

\begin{figure}
\centerline{\includegraphics[width=0.4\textwidth]{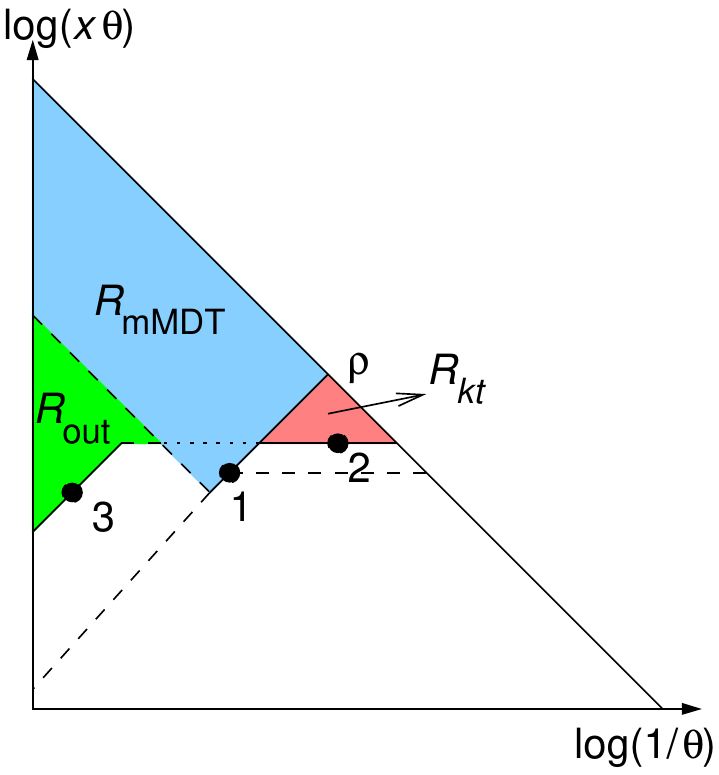}}
\caption{Representation of the various phase-space constraints and
  Sudakov exponents required for the resummation of the $\ln
  y$-enhanced terms for Y-splitter combined with the modified MassDrop
  tagger.}\label{fig:lund-mMDT-resum}
\end{figure}

In this Appendix we provide the details of the calculation leading to
Eq.~\eqref{eq:mmdt-resum} for a jet passing the Y-splitter condition and for which we study
the modified MassDrop mass. 
We work in the leading logarithmic accuracy and keep both leading
logarithms in $\rho$ and $\ycut$.

In this limit, we can assume that the groomed mass is dominated by a
single emission, say emission 1 with momentum fraction $x_1$ and at an
angle $\theta_1$ to the jet axis.
The fact that emission 1 is kept in the groomed jet guarantees that
$x_1>\ycut$.
We then have to consider four separate cases according to which emissions
dominate the $k_t$ and mass scales entering the Y-splitter
condition. 
We can write
\begin{align}\label{mmdt-resum-master}
\frac{\sigma}{\rho}\frac{d\sigma}{d\rho} = 
 & \int_y^1 dx_1\,P(x_1)\frac{\alpha_s(\rho x_1)}{2\pi} 
   e^{-R_{\text{mMDT}(\rho)}}\\
 & \bigg\{ e^{-R_{k_t}(\kappa_1;\rho)} \left[
     e^{-R_{\text{out}}(\rho;\kappa_1)}
     +\int_\rho^y\frac{d\rho_3}{\rho_3} R_{\text{out}}'(\rho_3;\kappa_1) e^{-R_{\text{out}}(\rho_3;\kappa_1)}
      \Theta(\kappa_1^2>y\rho_3)\right]\nonumber\\
 & + \int_{\kappa_{1}}^{\sqrt{\rho}} \frac{d\kappa_2}{\kappa_2} R_{k_t}'(\kappa_2;\rho)e^{-R_{k_t}(\kappa_2;\rho)}
   \bigg[ e^{-R_{\text{out}}(\rho;\kappa_2)} \Theta(\kappa_2^2>y\rho)\nonumber\\
 & \phantom{+ \int \frac{d\kappa_2}{\kappa_2} R_{k_t}'(\kappa_2;\rho)e^{-R_{k_t}(\kappa_2;\rho)}}
     +\int_\rho^y\frac{d\rho_3}{\rho_3} R_{\text{out}}'(\rho_3;\kappa_2) e^{-R_{\text{out}}(\rho_3;\kappa_2)}
      \Theta(\kappa_2^2>y\rho_3)\bigg]\bigg\}.\nonumber
\end{align}
In the above expression, the two terms on the second line correspond
to emission 1 also dominating the $k_t$ scale, while the last two
lines correspond to an additional emission 2 dominating the $k_t$
scale. In both cases, the plain jet mass can either be dominated by
emission 1 (the first term in each squared brackets) or by an
additional emission 3 (the second terms in each squared brackets).
Different terms are weighted by different Sudakov factors:
\begin{align}
 R_{\text{mMDT}}(\rho) &=
   \int\frac{d\theta^2}{\theta^2}dx\,P(x)\,\frac{\alpha_s(x^2\theta^2)}{2\pi}
   \Theta(x>y)\,\Theta(x\theta^2>\rho),\label{eq:sud-mmdt-Rmmdt}\\
 R_{k_t}(\kappa_i;\rho) &=
   \int\frac{d\theta^2}{\theta^2}dx\,P(x)\,\frac{\alpha_s(x^2\theta^2)}{2\pi}
   \Theta(x\theta>\kappa_i)\,\Theta(x\theta^2<\rho),\label{eq:sud-mmdt-Rkt}\\
 R_{\text{out}}(\rho;\kappa_i) &=
   \int\frac{d\theta^2}{\theta^2}dx\,P(x)\,\frac{\alpha_s(x^2\theta^2)}{2\pi}
   \Theta(x<y)\,\Theta(x\theta>\kappa_i\text{ or }x\theta^2>\rho).\label{eq:sud-mmdt-Rout}
\end{align}
These are graphically represented in Fig.~\ref{fig:lund-mMDT-resum}.
The $R_{\text{mMDT}}'(\rho)$, $R_{k_t}'(\kappa;\rho)$ and
$R_{\text{out}}'(\rho;\kappa)$ are the derivatives of the above
radiators wrt to the logarithm of (one over) their first argument.

\footnote{This
  corresponds to replacing $\Theta(x\theta^2>\rho)$ by
  $\rho\delta(x\theta^2-\rho)$ in (\ref{eq:sud-mmdt-Rmmdt}),
  $\Theta(x\theta>\kappa)$ by $\kappa \delta(x\theta-\kappa)$ in
  (\ref{eq:sud-mmdt-Rkt}), and
  $\Theta(x\theta>\kappa\text{ or }x\theta^2>\rho)$ by
  $\Theta(x\theta<\kappa) \rho\delta(x\theta^2-\rho)$ in
  (\ref{eq:sud-mmdt-Rout}).}
Note that the intermediate transition at $\kappa_i$ in
$R_{\text{out}}$ comes from the fact that an emission with $x<y$ and a
$k_t$ scale larger than $\kappa_i$ would dominate both the $k_t$ and
mass scales and the Y-splitter condition would not be satisfied. This
region is therefore automatically excluded.

Both integrations on $\rho_3$ can be performed quite
straightforwardly:
\begin{equation}\label{eq:mmdt-rho3-integration}
  \int_\rho^y\frac{d\rho_3}{\rho_3}
     R_{\text{out}}'(\rho_3;\kappa_i) e^{-R_{\text{out}}(\rho_3;\kappa_i)}
     \Theta(\rho_3<\kappa_i^2/y) =
     e^{-R_{\text{out}}(\kappa_i^2/y)}-e^{-R_{\text{out}}(\rho)}.
\end{equation}
In the above equation, we can drop the $\kappa$ argument of
$R_{\text{out}}(\rho;\kappa)$ for the following reason: for
$\rho<\kappa^2/y$, $x\theta>\kappa$ and $x<y$ automatically imply
$x\theta^2>\rho$ so that we can replace $\Theta(x\theta>\kappa_i\text{
  or }x\theta^2>\rho)$ by $\Theta(x\theta^2>\rho)$. We therefore have
\begin{equation}
 R_{\text{out}}(\rho) =
   \int\frac{d\theta^2}{\theta^2}dx\,P(x)\,\frac{\alpha_s(x^2\theta^2)}{2\pi}
   \Theta(x<y)\,\Theta(x\theta^2>\rho).
\end{equation}
Using \eqref{eq:mmdt-rho3-integration} for both squared brackets in
(\ref{mmdt-resum-master}), we obtain
\begin{align}\label{eq:mmdt-resum-base}
\frac{\sigma}{\rho}\frac{d\sigma}{d\rho} = 
 & \int_y^1 dx_1\,P(x_1)\frac{\alpha_s(\rho x_1)}{2\pi} 
   e^{-R_{\text{mMDT}(\rho)}}\nonumber\\
 & \left[e^{-R_{k_t}(\kappa_1;\rho)-R_{\text{out}}(\kappa_1^2/y)}
   + \int_{\kappa_1}^{\sqrt{\rho}} \frac{d\kappa_2}{\kappa_2}
   R_{k_t}'(\kappa_2;\rho)e^{-R_{k_t}(\kappa_2;\rho)-R_{\text{out}}(\kappa_2^2/y)}\right].
\end{align}

While this equation is suitable for practical purposes, specifically
numerical integration over $k_{t2}$ and $z_1$, it is not ideal to
see the logarithmic structure of the result. 
For that purpose it proves to be better to factor $\exp[-R_{\text{out}}(\rho)]$, which
would combine with the $\exp[-R_{\text{mMDT}}(\rho)]$ prefactor to
give the plain jet mass Sudakov, leading to (\ref{eq:mmdt-resum}).

\end{document}